\documentclass{mn2e}
\usepackage[dvips]{graphicx}
\usepackage{amsmath}
\usepackage{subfigure}
\usepackage{appendix}
\usepackage{natbib}
\usepackage{longtable}

\begin{document}

\title{A new hydrodynamics code for Type Ia Supernovae}

\author[S.-C. Leung et al.]
{S.-C. Leung,$^1$ M.-C. Chu,$^1$ L.-M. Lin$^1$ \\
$^1$Department of Physics and Institute 
of Theoretical Physics, The Chinese University 
of Hong Kong, Shatin, N.T., Hong Kong S.A.R., China}


\date{\today}

\maketitle
\begin{abstract}
A two-dimensional
hydrodynamics code for Type Ia supernovae (SNIa)
simulations is presented. The code includes a fifth-order
shock-capturing scheme WENO, detailed nuclear reaction
network, flame-capturing scheme and sub-grid turbulence.
For post-processing we have developed
a tracer particle scheme to record the thermodynamical
history of the fluid elements. We also
present a one-dimensional
radiative transfer code for computing observational signals. 
The code solves the Lagrangian hydrodynamics and 
moment-integrated
radiative transfer equations. A local ionization scheme
and composition dependent opacity are included. 
Various verification tests are presented, including
standard benchmark tests in one and 
two dimensions. SNIa models using the pure
turbulent deflagration model and the delayed-detonation
transition model are studied. The results are 
consistent with those in
the literature. We compute the detailed
chemical evolution using the tracer particles' histories,
and we construct corresponding bolometric
light curves from the hydrodynamics results. 
We also use a Graphics Processing Unit (GPU) to speed
up the computation of some highly repetitive
subroutines. We achieve an acceleration of
50 times for some subroutines and a 
factor of 6 in the global run time. 

\end{abstract}



\section{Introduction}
\label{sec:intro}

\subsection{Chandrasekhar Mass Explosion Model}

A Type Ia supernova (SNIa) is the explosion of 
a carbon-oxygen white dwarf (WD) due to 
thermonuclear runaway of carbon burning. 
It is believed to be a standard candle
due to its observed explosion homogeneity \citep{Branch1992}
and the luminosity-width relation \citep{Phillips1987}.
Also, in numerical modeling, the standard
Chandrasekhar mass WD is regarded as the progenitor
of an SNIa. These properties lead to wide applications
of SNIa as a cosmological ruler for determining
the Hubble parameters \citep{Leibundgut1992}, and the  
discovery of dark energy \citep{Riess1998, Perlmutter1999}.  

In the last few decades, several explosion 
mechanisms have been proposed, starting from the 
pure detonation model of Chandrasekhar mass
WD proposed in \cite{Arnett1969}. 
Despite its ability in unbinding the whole star,
the over-production of iron-peaked elements
and the absence of intermediate mass elements
(IME) make this scheme implausible \citep{Nomoto1977}. 
Later, the sub-sonic pure deflagration
model is proposed, which aims at providing 
sufficient time for electron capture in the 
laminar flame stage \citep{Nomoto1976} and production
of IME. The weakness of pure detonation model
is resolved but the deflagration
wave is too slow to unbind the WD.
The flame is quenched before 
consuming the whole WD, which leaves a large amount of 
unburnt material (fuel) \citep{Nomoto1976}. 
In view of this dilemma, several flame acceleration schemes
or transition schemes are proposed. Popular
models include the delayed-detonation transition 
model (DDT), the pure turbulent 
deflagration (PTD) model and the 
gravitationally confined detonation (GCD) model.

The DDT model is first suggested by Khokhlov (see for example
\cite{Khokhlov1989, Khokhlov1991a}). It is believed that the 
eddies around the flame front may provide
the required environment, which seeds
the detonation spot 
by the Zel'dovich gradient mechanism \citep{Khokhlov1991b}.
It has a counterpart in the shocktube experiment. 
The model has been found satisfactory
because of the sufficient production of IME,
absence of remnant and consumption of 
fuel around the core \citep{Khokhlov1989, Gamezo2004, Gamezo2005}.
However, whether the first detonation spot
can be seeded is still an open question. 
It is believed that the 
velocity fluctuations are adequate for 
triggering the detonation \citep{Khokhlov1995}. 
On the other hand, several turbulent flame 
simulations (see for example \cite{Niemeyer1999a}, \cite{Imshenik1999}, 
and \cite{Lisewski2000}) have
shown that the developed velocity fluctuations 
are far below the required level. Also, the detonation
front is found to be unstable when encountering
obstacles \citep{Maier2006}. This makes
the robustness of this mechanism in doubt.

The PTD model assumes that the 
the fluid is highly turbulent
in view of the high Reynolds number \citep{Niemeyer1995b}. 
Local velocity fluctuations are 
important for the flame evolution because
the flame is constantly stretched and 
perturbed by the fluid motion, which 
enlarges the local burning surface \citep{Khokhlov1994}. 
This increases the fuel consumption rate
and hence makes the flame propagate
faster, compared with a laminar flame 
under the same condition \citep{Timmes1992b}. 
Two models are commonly considered in 
turbulent flame modeling. 
The first model is proposed in \citep{Damkoehler1939}.
It is observed in Bunsen flame 
experiments that the turbulent flame 
propagation speed $v_{{\rm turb}}$ 
varies with local turbulence fluctuations $\tilde{v}$:
$v_{{\rm turb}} = \tilde{v}$ \citep{Niemeyer1995a}. The
second model is obtained from 
a theoretical analysis \citep{Pocheau1994}.
It is found that $v_{{\rm turb}} = v_{{\rm lam}} 
\sqrt{1 + C (\tilde{v} /v_{{\rm lam}})^2}$
with $v_{{\rm lam}}$ being the laminar flame speed. 
The applications of turbulent flame 
in the SNIa explosion scenario are first
proposed in \cite{Niemeyer1995a}, based on the sub-grid turbulence
model reported in \cite{Clement1993}. The model
has been found successful in providing a 
healthy explosion in two-dimensional 
\citep{Reinecke2002a} and three-dimensional simulations
\citep{Reinecke2002b, Roepke2005a, Roepke2005b}. 
However, several drawbacks have been found in 
detailed three dimensional studies.
They include the presence of low-velocity
fuels (carbon and oxygen) in the ejecta
and underproduction of IME \citep{Roepke2007b}.

The GCD model, also known as the 
Deflagration Failed Detonation (DFD) 
model \citep{Plewa2007, Kasen2007},
is similar to the DDT model,
which starts from the deflagration phase.
But the burning is weak so that the star 
remains bound. The hot and burnt material 
causes the WD to expand, floating
the hot flame by buoyancy
to the surface. After then, a large 
amount of fuel remains unburnt. 
The flame propagates throughout the WD
by a surface flow \citep{Jorden2008}. The flow finally
merges at one point, usually at the point opposite
to the breakout location \citep{Meakin2009}. The 
converging flow heats up the cold and
low-density fuel, pushing it deep inside
the star. Once the squeezed fuel becomes the hot spot
for detonation as its density exceeds the threshold, 
a detonation front forms, which burns the remaining stellar
material and unbinds the whole WD \citep{Jorden2012}. 
The qualitative difference between DDT model and GCD model
is that, in the former, the detonation starts
from the inside of the WD and sweeps outward, while it is 
the opposite in the latter. 

Recent developments of hydrodynamics
models have considered extensively all three 
models. In \cite{Long2014}, three-dimensional 
PTD models are studied. In \cite{Seitenzahl2013}, 
the DDT model is shown to be able to explain certain SNIa
observation data. In \cite{Blondin2011} and 
\cite{Blondin2012}, DDT models in one-dimension
and two-dimension are studied with their synthetic
spectra and light curves. Such fine
details provide a mean to constrain the explosion
model \citep{Dessart2013}. 

\subsection{Radiative Transfer}

The modeling of post-explosion
light curve and spectrum are crucial for 
discriminating the validity of an explosion 
mechanism. The governing physics of SNIa
light curves are believed to be the decay 
of synthesized $^{56}$Ni into $^{56}$Co
in early time, and decay of $^{56}$Co 
into $^{56}$Fe in late time \citep{Colgate1969}. 
Simplified models, including those with
diffusion approximation \citep{Colgate1980}, or analytic models 
assuming a fireball undergoing homologous 
expansion \citep{Arnett1982}, can already capture primary 
features of SNIa light curves and match a number of 
observed SNIa. 

Despite that the light curves are well fitted by these models, in order to 
understand the variety of SNIa observations,
as well as their relations with explosion models,
detailed radiative transfer for both 
bolometric and multiple wavebands are important \citep{Zhang1994}.
The actual problem of radiative transfer
can be highly nontrivial due to its interactions
with hydrodyanmics, the presence of millions to 
billions of atomic transition lines, ionization
and excitation of atoms, and the differential-integral 
structure of the radiative transfer equations. 
These features have been studied in details
in the last few decades. Physics and numerical factors, such as the  
effects of line blanketing \citep{Hillier1990a, Hillier1998}, 
consistent boundary conditions \citep{Sauer2006},
treatment in opacities \citep{Hoeflich1993}, 
acceleration techniques \citep{Hillier1990b, Lucy2001}
and so on have been studied extensively. 
A number of numerical codes for solving this 
problem have been under constant
development in the past few decades 
(see STELLA \citep{Blinnikov1998, Blinnikov2000, Blinnikov2006}, 
ARTIS \citep{Lucy2005, Sim2007a, Kromer2009},
SEDONA \citep{Kasen2006} 
and PHOENIX \citep{Hauschildt2010, Seelmann2010, Hauschildt2011} 
for the instrument papers and realizations).

Due to the stringent demand in computational resource 
for calculating radiative transfer, in previous studies, the explosion 
phase and its observational consequences are usually modeled
separately \citep{Nomoto1986, Zhang1994, Hoeflich1995,
Nugent1997}, where the radiative transfer part makes use
of the hydrodyanmics results from some benchmark runs.
Recent development of computational power 
allows the spectral synthesis to be coupled 
in the hydrodynamics to form a pipeline following the 
explosion and homologous expansion 
phases. Different explosion mechanisms
have been studied with fine details, such as 
the PTD model \citep{Fink2014, Long2014}, 
DDT model \citep{Blondin2011, Seitenzahl2013} and
GCD model \citep{Plewa2007, Kasen2007}.

\subsection{Flame Capturing Scheme}

The flame capturing scheme is a technique of describing
discontinuities with details in sub-grid scales.
Such discontinuities can be in forms of fluid 
type, composition or characters. In SNIa simulations,
it is essential because the typical width of 
deflagration waves is in the 
order of centimeters or even smaller. Also, their
propagation speed is much slower than the fluid
speed of sound. Therefore, within one time
step under the Courant-Fredrich-Lewy 
condition, only a partial amount of a 
fluid element in a grid is 
burnt and so it is important to determine
the ash-fuel interface in partially 
burnt grids.

To account for such discontinuities, three types 
of interface tracking algorithms
are commonly used. The first one is the level-set 
method \citep{Osher1988}. This method suggests that
the discontinuity can be described by the zero
contour of a scalar function, which is interpreted
as the distance function. 
This method has advantages in its easy implementation
and direct coupling with hydrodynamics. Also, 
the scheme can be generalized directly to arbitrary
dimensions. The topological changes are handled
naturally without the need of extra spotting
mechanism \citep{Sethian2001}. However, it is also known to have
poor volume conservation \citep{Rider1995}. The distance function
requires reinitialization at each iteration
step in order to preserve the distance function 
validity \citep{Sussman1994}. This
scheme in SNIa is first proposed by \cite{Reinecke1999a},
and applications are found in \cite{Reinecke1999b, Reinecke2002a, Reinecke2002b}. 

Another algorithm is the point-set method \citep{Glimm1981, Glimm1985, Glimm1988}. 
The flame front is described as 
a line in a two-dimensional simulation
or a surface in a three-dimensional simulation, 
by a linked set of massless particles which 
are advected along the streamline of fluid flow, and 
they only reveal the position of the interface
without affecting the hydrodynamics properties of the
fluid. This method receives wide applications in 
other fields, such as the study of bubbles \citep{Youngren1976}.
It has a huge advantage that the interface
properties are exactly modeled, such that their
geometric properties can be obtained in a 
straightforward manner \citep{Tryggvason2001}. However, it has two major 
shortcomings. First, it requires special
attention in topological changes, 
namely surface splitting and merging \citep{Glimm1988}. 
Also, the extension of the point-set method 
to three-dimensional simulations is 
non-trivial, because the surface is no longer 
represented by a line segment \citep{Glimm1999}. For example of application, 
see \cite{Zhang2009} for a laminar deflagration
model. It is shown that, with sufficient fine details,
the laminar flame can bring a successful explosion without
the need of any flame acceleration scheme. 

The third mechanism which is commonly found in the 
literature is the volume of fluid method \citep{Hirt1981}. 
This method is similar to the 
level-set method, which introduces an extra 
scalar field that represents the volume
fraction of a specific fluid. This model
is also regarded as the advection-diffusion-reaction
equation \citep{Calder2007, Townsley2007}. 
The method is known for its exact 
conservation of mass \citep{Scardivelli1999}. 
However, the implicit nature of geometric quantities
becomes its disadvantage.
The geometry of the flame, which is important in reconstructing
the thermodynamics of partially burnt grids,
is dependent of the reconstruction scheme \citep{Rudman1997}.
In SNIa context, the algorithm is pioneered 
in simulations presented in \cite{Khokhlov1993b}. The FLASH
code has adopted this algorithm as the default
flame capturing scheme; see for example \citep{Calder2007}. 

\subsection{Sub-grid Turbulence}

The fluid motion can be turbulent 
because of the high Reynolds number.
The flame surface is known to be unstable
subject to hydrodynamic instabilities related to
turbulence. The eddies stretch and perturb the flame front,
which increase its effective burning area
and enhance burning rate \citep{Timmes1992b}, when compared with 
laminar flame at the same density. Similar to the difficulties 
in resolving the flame interface, the eddies have sizes 
down to the Kolmogorov's scale ($\sim$ cm) \citep{Woosley2009}, 
and therefore a direct numerical modeling of 
these eddies is not yet feasible. Instead, sub-grid turbulence
is employed to mimic the small-scale 
eddies by scaling kinematics from large scales. 

Various implementations
of sub-grid turbulence models are proposed. 
They can be classified mainly into the
one-equation model, two-equation
model and Reynolds stress model. In all these models, 
the production/dissipation rates are closed by
direct numerical simulations, statistical closure
or dynamical modeling.

The one-equation model, first proposed by 
\cite{Smagorinsky1963}, introduces the turbulence
kinetic energy $q$. It first appears in the SNIa simulation in 
\cite{Niemeyer1995a}, which relies on a statistical
model based on Kolmogorov's scaling 
\citep{Clement1993}. The model gives
breakthroughs by providing a healthy explosion, as
shown in \cite{Reinecke2002a}. However, it is argued
that the dissipation might not be following
Kolmogorov's one instantaneously \citep{Schmidt2006b}. 
The model is later extended in \cite{Schmidt2005}, 
which generalizes the eddy production or dissipation 
terms dynamically. 

The two-equation model, first proposed 
in \cite{Launder1974}, introduces two extra quantities, 
which model the decay of eddies by
an extra dissipation term $\epsilon$ ($q-\epsilon$
model) or a vorticity term $\omega$ ($q-\omega$ model).
Both models aim at describing the dissipation of 
eddies. However, the original equations do not 
guarantee the positivity of $q$ and $\epsilon$ 
(non-realizable). We refer the reader to \cite{Shih1994}
for modifications of the equations that can 
guarantee the realizability condition.
Unlike the one-equation 
model, this model is not applied 
in common SNIa simulations. 

The Reynolds stress model provides algebraic closure
to the sub-grid velocity correlations \citep{Shih1995}. 
However, not all the variables can be 
closed by direct numerical simulations due to the larger number of 
coefficients. A similar problem can already be found
in some extensions of the two-equation model.
For instance, in the three-equation model, an 
ad hoc timescale is needed in order to 
close the equation set \citep{Yoshizawa2012}.

\subsection{Motivation and Structure}

In SNIa literature several codes are commonly used for 
studying the explosion phase. The first one is 
LEAFS, which is known
as PROMETHEUS in earlier publications (see
for example \citep{Reinecke1999b}).
This code makes use of the piecewise-parabolic method (PPM) \citep{Collela1984}
with moving meshes that follow the WD expansion
(see \cite{Roepke2005a, Roepke2005b} for the 
development and early applications). 
The code models the turbulent nuclear
flame using a sub-grid turbulence algorithm
with the level-set method as flame-tracking scheme.  
This code has been used in large-scale massively parallel
runs in PTD \citep{Fink2014} and DDT models \citep{Seitenzahl2013}. 
Another hydrodynamics 
code is the FLASH code \citep{Fryxell2000}.
This code is embedded with adaptive-mesh refinement
such that the flame structure can be traced
with fine details.  
Recently, FLASH has been developed into a 
part of a pipeline in SNIa numerical tool 
in constructing, analyzing and comparing 
numerical data with observations \citep{Long2014}.
The code uses the advection-diffusion-reaction
scheme in describing the ash distribution and
flame geometry but without sub-grid turbulence. 
Recent large-scale computations in the PTD
models \citep{Long2014} and GCD models \citep{Jorden2012}
have used this open-source code. 
We compare our code with LEAFS and FLASH in Table 
\ref{table:CodeStructure}. 

There are two motivations for us to build our own 
hydrodynamics code. First, at the time when the 
code was being developed, there was not yet
a published SNIa code which makes use
of a 5th order scheme in spatial discretization.
Second, we shall apply the hydrodynamics code 
to various astrophysical contexts. One of our
work in progress is to study the effects 
of dark matter on SNIa explosions 
\citep{Leung2015b}. By building our own code, it is 
easier for us to implement and 
analyze any additional physics component
in SNIa simulations.

\begin{table*}
\begin{center}
\caption{Comparison of the numerical components and 
structure of our code with other commonly used hydrodynamics codes.}
\begin{tabular}{|c|c|c|c|}
\hline
Scheme & Our code & LEAFS (PROMETHEUS) & FLASH \\ \hline
Spatial discretization & WENO 5$^{{\rm th}}$ order & PPM & PPM/WENO \\
Accuracy & $5^{{\rm th}}$ order & $3^{{\rm rd}}$ order & $3^{{\rm rd}}-5^{{\rm th}}$ order \\ \hline
Time discretization & Runge-Kutta 5-Step & Directional splitting & Directional splitting \\
Accuracy & $3^{{\rm rd}}$ order & $2^{{\rm nd}}$ order & $2^{{\rm nd}}$ order \\ \hline
Adaptive Mesh Refinement & No & No & Yes \\
Dimensions & 1-2 & 1-3 & 1-3 \\ \hline
Flame capturing scheme & Level-set ($default$) & Level-set & ADR \\ 
 & Point-set & & \\
Sub-grid turbulence & Yes & Yes & No \\
Number of online isotopes & 19 & 5 & 3 \\ \hline
Hardware acceleration & GPU & Multi-processors & Multi-processors \\ \hline

\end{tabular}
\label{table:CodeStructure}
\end{center}
\end{table*}

In this article, we report 
numerical tests of our hydrodynamics code and 
comparisons with other
benchmark tests reported in the literature. 
In Section \ref{sec:methods}, we describe 
the hydrodynamics equations and the numerical
techniques that we implement in our code.
In Section \ref{sec:physics} we 
describe the input physics for modeling SNIa. We also
introduce the radiative transfer code by 
describing its numerical implementation.
Then in Section \ref{sec:post} we outline
the post-processing tools we have used in 
extracting observables from the 
hydrodynamics results, including the 
tracer particle method and the one-dimensional 
moment-integrated radiative transfer code.  
In Section \ref{sec:results}, we report our results.
We describe the configurations and results of
some benchmark one- and two- dimensional
tests. We also show two test runs using the 
PTD and DDT models, and we compare our results
with some similar runs in the literature. 
In Section \ref{sec:obs} we describe the 
results after post-processing, including
the detailed nucleosynthesis yields, bolometric
light curves and the synthetic spectra. 
In Section \ref{sec:discussion},
we summarize and briefly outline 
our future plans. 
 
\section{Methods}
\label{sec:methods}

\subsection{Hydrodynamics}
We have developed a two-dimensional Eulerian hydrodynamical 
code in cylindrical coordinates to model SNIa. 
The equations include
\begin{eqnarray}
\frac{\partial \rho}{\partial t} + \frac{\partial \rho v_r}{\partial r} 
+ \frac{\partial \rho v_z}{\partial z} = - \frac{\rho v_r}{r}, \\
\frac{\partial \rho v_r}{\partial t} + \frac{\partial (\rho v_r^2 + p)}{\partial r} 
+ \frac{\partial \rho v_r v_z}{\partial z} = 
- \frac{\rho v_r^2}{r} - \rho \frac{\partial \Phi}{\partial r}, \\
\frac{\partial \rho v_z}{\partial t} + \frac{\partial \rho v_r v_z}{\partial r} 
+ \frac{\partial (\rho v_z^2 + p)}{\partial z}= -\frac{\rho v_r v_z}{r}
- \rho \frac{\partial \Phi}{\partial z}, \\
\frac{\partial \tau}{\partial t} + \frac{\partial (\tau + p) v_r}{\partial r} 
+ \frac{\partial (\tau + p) v_z}{\partial z} = \notag \\
- \frac{(\tau + p) v_r}{r} - \rho \left( v_r \frac{\partial \Phi}{\partial r} + 
v_z \frac{\partial \Phi}{\partial z} \right) + \dot{Q}. 
\label{eq:euler2D}
\end{eqnarray}
In the above equations, $(r,z)$ are
the distances from the polar axis and the symmetry plane. 
$\rho$, $v_r$, $v_z$, $p$ and $\tau$ are the mass
density, velocities in $r$ and $z$ directions, pressure 
and total energy density of the fluid. The total energy
density includes both the thermal and kinetic
parts $\tau = \rho \epsilon + \frac{1}{2} \rho v^2$, 
where $\epsilon$ is the specific internal energy. $\Phi$, the 
gravitational potential, satisfies the Poisson Equation 
$\nabla^2 \Phi = 4 \pi G \rho$.   
$\dot{Q} = \dot{Q}_{{\rm nuc}} - \dot{Q}_{\nu} -
\dot{Q}_{{\rm turb}}$ is the heat source and loss 
from nuclear fusion, neutrino emission and 
sub-grid turbulence. Note that we deliberately 
arrange the Euler equations
into this conservative form in order to couple them with 
the Weighted Essential Non-Oscillatory scheme (WENO)
directly. The terms on the right hand side 
are regarded as source terms. 

We use a high-resolution shock-capturing scheme, 
WENO \citep{Barth1999}, for spatial discretization. 
This is a fifth-order scheme, which processes piecewise 
smooth functions with discontinuities in order to 
simulate the flux across grid cells with high precision, 
while avoiding spurious oscillations around the shock. 
In the code, we use a finite-difference formulation 
of the WENO scheme while the hydro quantities 
in the simulations, such as $\rho$, $v$, $\epsilon$
and so on, represent the averaged values of the 
actual profiles across the Eulerian grid. For the 
construction of numerical fluxes, we use the 
WENO-Roe scheme \citep{Jiang1996}. 
First we separate the 
hydro flux into a positive flux and a negative flux 
by a global Lax-Friedrichs splitting. Then we apply 
the WENO reconstruction on the two fluxes by an 
upwinding scheme. The numerical flux is 
obtained by summing the reconstructed 
positive and negative fluxes.

For the discretization in time, we use 
the five-stage, third-order, non-strong-stability-preserving 
explicit Runge Kutta scheme \citep{Barth1999}.
Note that in the original prescription the 
three-stage, third-order, strong-stability-preserving
explicit Runge-Kutta scheme is suggested. However,
comparison of these two schemes shows that the latter one 
has no stability advantage and is less efficient
\citep{Wang2007}. Therefore, we adopt the new
time-discretization scheme. 

The gravitational potential $\Phi$ is obtained by solving
the Poission equation $\nabla^2 \Phi = 4 \pi G \rho$.
We use the successive over-relaxation method with the
Gauss-Seidel method to solve the Poisson equation in 
cylindrical coordinates. The iteration is stopped by 
default when the relative changes of all grids reach
below 10$^{-8}$. The inner boundaries are assumed to be 
reflecting and the outer boundaries are fixed by using 
the Roche approximation, which is given 
by $\Phi(r,z) = -GM/ \sqrt{r^2 + z^2}$. 
We remark that the Roche approximation assumes
point-like mass everywhere, but in general 
higher multipole terms are needed. In our case, since the outer
boundaries lie about a few times of the 
initial stellar radius away from the 
center of the star, the higher multipoles' 
contributions are negligible. We find that 
the quadruple term gives a less than $1 \%$
correction to the monopole term 
throughout the simulations.

\subsection{Level-set method}
\label{sec:methods_LSM}

We use the level-set method to trace the geometry of the
flame front. This method is widely applied to other areas
which require a detailed description of surfaces. It has been
applied in SNIa modeling as prescribed in details 
in \cite{Reinecke1999a}.
For the sake of completeness, we outline the governing
equations and its procedure. 

The level-set method introduces an extra scalar
field $S$ imposed on an Eulerian grid, which 
satisfies $|\nabla S| = 1$. The flame front is 
defined by the zero-contour of the field. We
also define grids with positive (negative)
values of $S$ to be full of ash (fuel). 
Geometrically speaking, the value is
the minimum separation of a given point 
to the zero-contour surface. 

The flame front propagates by both fluid advection and 
flame burning, which can be written as
\begin{equation}
\frac{\partial S}{\partial t} = -( \vec{v} + \vec{v}_{{\rm flame}}) \cdot \nabla S.
\end{equation}
$\vec{v}$ is the velocity of the fluid, governed by 
the compressible Euler equations. $\vec{v}_{{\rm flame}}$
is the effective flame propagation speed 
obtained from the nuclear reactions  
described in Sec. \ref{sec:Network},
which points towards the outward normal direction
of the flame surface. For the dependence on density
or composition of the flame speed, see in Appendix \ref{sec:defwave} 
their derivations and analytic fits.

With the evolution equations, we may simultaneously
evolve the flame front coupled with the hydrodynamics.
Note that in our simulations, the scalar field is defined on the 
corners of a grid, such that the position of 
flame front on the grid boundary can be found exactly. 

In updating the scalar field, we make use of 
operator splitting to deal with the contributions
from fluid advection and flame propagation. The 
latter part can be done with a common 
high-order advection scheme such as the 
piecewise-parabolic method (PPM) \citep{Collela1984}
or the WENO scheme, such that the advection
may preserve the area (volume) of the flame. 
Naturally, we use the WENO scheme as we have 
used to solve the hydrodynamics. 
After each update in $S$, we need to do a
reinitialization (or redistancing) such that the 
scalar field $S$ can maintain the distance property
$|\nabla S| = 1$. We describe the procedure below. 

First, we enumerate all positions where the flame front
cut the grid, by finding all 
neighboring pairs of the scalar field where its
values are of opposite signs. The position
of the flame front is obtained by linear interpolation. 
Then the distances of all grid points to the flame
front positions are compared, and the smallest one is chosen
as the new value of $S$.

Note that this procedure preserves the distance property of 
$S$, but it perturbs the current flame position as well.
Therefore, a smooth transition from its original 
value to the corrected value is used. When the grid is 
far away from the flame front, the distance value is used,
otherwise the original value is kept. Mathematically, for 
a given scalar field $S_{{\rm old}(i,j)}$ with a
minimum flame distance $d_{(i,j)}$, we have  
\begin{equation}
S_{{\rm new}(i,j)} = H(d_{(i,j)}) S_{{\rm old}(i,j)} + (1 - H(d_{(i,j)})) d_{(i,j)},
\end{equation}
where $H(d)$ is a smooth function satisfying 
$H(0) = 1$ and $H(\infty) = 0$. In our case,
we follow the choice of \cite{Reinecke1999b} that
\begin{equation}
H(d) =  \frac{1 - {\rm tanh} \frac{d - d_0}{\Delta/3}}{1 - {\rm tanh} \frac{- d_0}{\Delta/3}},
\end{equation}
with $d_0 = 3 \Delta$, $\Delta$ being the 
grid size in our simulations. This formula implies
that the values of $S$ on the three closest neighboring
grids around the flame front are unchanged.

As we have mentioned, the level-set method is 
not the only flame-capturing method. We have also 
developed the point-set method in order to 
compare their performances. See Appendix \ref{sec:methods_PSM}
for the details and numerical results. 

\subsection{GPU accelerations}

In view of a large number of subroutines involved
in each time step, we share parts of the 
computation with a graphics
processing unit (GPU). By allowing the 
computations to be simultaneously done 
by more than thousands of threads, 
the total computation time can be
drastically reduced, especially when the job 
requires repetitive but similar operations.

One major difference in CPU coding and GPU coding is
the lack of static memory in GPU. Such a feature has
made the use of GPU in typical hydrodynamics 
simulations unfavorable because the static memory 
provides convenient access to data, including
the hydrodynamics variables and chemical
composition, without the need of passing 
a large amount of variables among subroutines.
Furthermore, such data migration, if not handled
carefully, can take up even more time than
using CPU solely. Nevertheless,
the use of GPU can be advantageous when applied to
subroutines that do not have
the above undesirable properties.

In Table \ref{table:TestGPU} we tabulate the 
subroutines being processed by an NVIDIA GT
Titan Black GPU and their running time, 
compared with an Intel I7 quad-core CPU. 
They include 
the Poisson solver, reinitialization
procedure in the level-set method, WENO 
scheme and spatial discretization in other subroutines.

In general, two classes of time reduction can be 
found. For subroutines with pure matrix operations
or repetitive procedures, such as the Poisson
solver or the reinitialization scheme, 
a factor of 20-40 reduction can be obtained. 
For subroutines with heavy data transfer
or significant amount of logical operations,
such as the WENO scheme or the spatial
discretization (which includes flux limiter),
the time reduction is less significant, about a
factor of 2 or less. Overall, there is a factor of 6
reduction in run time by using both CPU and GPU, compared with 
runs which make use of CPU only.

\begin{table}
\begin{center}
\caption{A sample run of the code and a 
comparison of running time of subroutines 
with GPU version in one step. Subroutine time lapses
are in unit of seconds and the full run is in 
units of hours. CPU: Intel I7, GPU: 
NVIDIA Titan Black.}

\begin{tabular}{|c|c|c||}
\hline
subroutines & run time & run time \\
 & CPU only & CPU + GPU \\ \hline
WENO scheme (s) & 3.45 & 2.06 \\
Poisson solver (s) & 468.42 & 12.07 \\
Reinitialization (s) & 20.18 & 1.20 \\ 
Spatial discretization (s) & 1.08 & 1.10 \\ \hline
One full run (hour) & 199.66 & 29.78 \\ \hline
\label{table:TestGPU}
\end{tabular}
\end{center}
\end{table}

\section{Input physics}
\label{sec:physics}

\subsection{Equation of states}

To close the equation set, we use the open-source 
Helmholtz equation of state (EOS)
reported in \cite{Timmes1999a}
and \cite{Timmes2000a}.
The EOS describes the equilibrium thermodynamical
properties of a gas, which includes 1. electrons in the form of ideal 
gas with arbitrarily degenerate and relativistic levels, 2. ions
in the form of classical ideal gas, 3. photons in Planck distribution, and
4. contributions from electron-positron pairs. 
The subroutine takes input of $\rho$, temperature $T$,
mean atomic mass $\bar{A}$, mean atomic number $\bar{Z}$ and gives other
thermodynamical quantities, including $p$, $\epsilon$, adiabatic
index $\gamma$ and so on, together with their derivatives
with respect to each input parameter. 

\subsection{Nuclear Reaction Network}
\label{sec:Network}

To calculate the nuclear reaction heat production $\dot{Q}_{{\rm nuc}}$,
we use the 19-isotope nuclear reaction network
subroutine reported in \cite{Timmes1999b}. 
The isotopes include $^{1}$H, $^{3}$He, $^{4}$He, 
$^{12}$C, $^{14}$N, $^{16}$O, $^{20}$Ne, $^{24}$Mg, 
$^{28}$Si, $^{32}$S, $^{36}$Ar, $^{40}$Ca, $^{44}$T,
$^{48}$Cr, $^{52}$Fe, $^{54}$Fe, $^{56}$Ni, neutron
and proton. The fusion network includes series of 
reactions from hydrogen burning
up to silicon burning. Both $(\alpha,\gamma)$
and $(\alpha, p)(p, \gamma)$ reactions 
are included. Other isotopes, including 
$^{27}$Al, $^{31}$P, $^{35}$Cl and so on are included implicitly. 
Certain isotopes are less important here, for example, 
$^{1}$H, $^{3}$He and $^{14}$N. They are involved in 
hydrogen burning or CNO-cycle, which are not crucial
in SNIa. On the other hand, isotopes along the $\alpha$-particle chain 
are important for the exothermic processes in the explosions,
as most energy is released by these processes, and the amount
of $^{56}$Ni is important for observations. 

We do not couple the nuclear fusion subroutine
directly in the code except at low density for two reasons.
First, the deflagration and detonation wave fronts 
at high density have widths much thinner than the numerical grid size,
and so only a fractional change occurs in the 
grids which are partially burnt in each time step. The energy release and isotope
variations will be over-estimated if the whole cell is 
considered. Hence, the energy release relies on 
the fractional changes of area (volume) being consumed 
by the deflagration and detonation wave front,
which depends strongly on the local density, but weakly on 
the local temperature. On the other hand, at low density, 
the deflagration and detonation wave can have a width 
comparable with simulation grid size. Then computing the nuclear
reaction yields with data from the whole cell is 
acceptable. Second, the accuracy cannot be much improved
unless the isotope transport of the burning grids
are well described. But an exact reconstruction 
is time-consuming and sensitive to the numerical 
accuracy. As shown in \cite{Reinecke1999a}, an 
exact reconstruction is difficult for grids with 
large numerical errors. As a result, we choose the less
accurate scheme by using tables of reaction products 
and energy release as functions of the input density.
We describe the construction method and 
results in Appendix \ref{sec:defwave} and Appendix \ref{sec:detwave}.

\subsection{Sub-grid turbulence}

The use of the Euler equations is a good approximation when
the fluid viscosity is small. This is true for 
the case of a WD. In a WD the fluid has a typical 
Reynolds number $R_e \sim 10^{14}$.
This gives a dissipation range $\eta$ related to 
the integral range $L$ by the Reynolds number 
\begin{equation}
L = \eta R_e^{3/4}.
\end{equation}
Therefore, the dissipation range is about $10^{-3}$ cm, 
and the scales in which turbulence needs
to be directly modeled are far smaller than the 
simulation grid size $\Delta \approx 10^6$ cm.
In this range $\eta < \Delta < L$, called the inertial range, 
the hydrodynamics is dominated by the 
inertial term in the Navier-Stoke Equation and the 
turbulent dissipation is independent of small-scale
viscosity. This
allows us to treat the fluid as an ideal fluid, and take 
the effects of unresolved turbulence statistically. 
In particular, the turbulent velocity satisfies
the Kolmogorov's scaling law, which suggests that the local
velocity fluctuations $\tilde{v}(l)$ are related to 
the local length scale $l$ by 
\begin{equation}
\tilde{v}(l) = \tilde{v}(L) \left( \frac{l}{L} \right)^{1/3},
\end{equation}
and there is a constant energy transfer rate from large to small scales.

The sub-grid turbulence model is based on the above 
properties as described in \cite{Clement1993}. This
model is originally used to resolve the problem of
frozen flow as commonly found in simulations of astrophysical
convection. Such flow consists of unphysically large scale 
flow in opposite directions within neighbouring cells.
Note that this result is also a solution of the 
Euler equations. It appears because the simulation fails to 
recognize that such large velocity gradient can actually
result in sub-grid turbulence, which dissipates the excess
velocity gradient into thermal energy by its kinematic viscosity. 
In practice, the effects of sub-grid turbulence
are included as a heat source $\dot{Q}_{{\rm turb}}$
\begin{eqnarray}
\dot{Q}_{{\rm turb}} = -A \rho q \nabla \cdot v + \Sigma_{ij} \frac{\partial v_i}{\partial x_j}  
- \rho \epsilon_{dis} + C_{{\rm Arch}} \rho g_{{\rm eff}},
\label{eq:turb_eq}
\end{eqnarray}
where 
\begin{equation}
\Sigma_{ij} = \rho \nu_{{\rm turb}} \left( \frac{\partial v_i}{\partial x_j} + 
\frac{\partial v_j}{\partial x_i} - A \nabla \cdot v \delta_{ij} \right)
\end{equation}
is the shear-stress tensor of the fluid flow. 
$C_{{\rm Arch}}$ is chosen such that the flame propagation speed
recovers the limit when gravity is important and Rayleigh-Taylor 
instabilities become the major flame acceleration mechanism. 
We have $C_{{\rm Arch}} \approx 1/2$ in our case. 
$A$ is a dimensionless constant with $A = 1$ ($2/3$) for 
2- (3-) dimensional fluid flow \citep{Niemeyer1995a, Reinecke2002a}. 
The terms on the right hand side of Eq. (\ref{eq:turb_eq})
are regarded as sources, which include turbulence production 
by compression and shear, turbulence dissipation and Archimedian production.
$g_{{\rm eff}}$ is the effective
gravitational force defined by the product of the 
local gravitational force and the Atwood number.
The Atwood number characterizes the density 
contrast between burnt and unburnt matter,
which are functions of density, as presented 
in \cite{Timmes1992a}. $q$ is the specific turbulence kinetic 
energy $q = \tilde{v}^2/2$, which evolves as a 
scalar field and couples to the fluid through $\dot{Q}_{{\rm turb}}$ by
\begin{eqnarray}
\frac{\partial \rho q}{\partial t} + \nabla \cdot \left( \rho \vec{v} q \right) = 
\dot{Q}_{{\rm turb}} + \nabla \cdot (\rho \nu_{{\rm turb}} \nabla q),
\label{eq:energy_eq2}
\end{eqnarray}
where the last term on the right hand side
corresponds to turbulence diffusion. 

In this paper, in order to compare consistently
with results from other hydrodynamics codes, we use
the algebraic closure proposed in \cite{Khokhlov1995}. The 
turbulence generation/dissipation terms are derived
from Kolmogorov's scaling relation, namely
\begin{equation}
\nu_{{\rm turb}} \approx \Delta \tilde{v} (\Delta) = C \Delta q^{1/2},
\label{eq:turbnu}
\end{equation}
and 
\begin{equation}
\epsilon_{{\rm dis}} \approx \frac{\tilde{v}^3}{\Delta} = D \frac{q^{3/2}}{\Delta},
\label{eq:turbeps}
\end{equation}
with $C$ and $D$ as parameters.
In \citep{Clement1993} $C$ and $D$
are modeled in analogy to the "wall proximity functions"
by defining the parameter $W = \epsilon / q$. It is found that
\begin{equation}
C = 0.1 F,
\label{eq:turbC}
\end{equation} 
\begin{equation}
D = 0.5 / F,
\label{eq:turbD}
\end{equation}
with 
\begin{equation}
F = \min [100, \max (0.1, {10^{-4}} W)].
\label{eq:turbF}
\end{equation}
The term $F$ ensures that the generation term is 
large when $q$ is small (laminar flow) and 
the dissipation term is large when $q$ is large (turbulent flow). 
We remark that even we have mentioned Kolmogorov
scaling at the beginning of this section, this sub-grid turbulence
model does not assume any particular scaling relation, since
the forms of Eqs. (\ref{eq:turb_eq}) and (\ref{eq:turbeps})
are obtained by only dimensional analysis. 
Individual scaling relations in the fluid flow can be 
realized by carefully tuning Eqs. (\ref{eq:turbC}) - (\ref{eq:turbF}).
But as indicated by \cite{Clement1993}, this procedure
can be very difficult unless one solves the spectral equations
simultaneously, which is unfeasible in the current stage.
Nevertheless, the current form is shown in the literature 
that it can describe astrophysical flow with a high 
Reynolds number well.

Following the suggestion of \cite{Schmidt2006b}, the turbulent
flame propagation speed is connected to the turbulent
velocity fluctuation by
\begin{equation}
v_{{\rm flame}} = v_{{\rm lam}} \sqrt{1 + 2 C_t \left( \frac{q}{v_{{\rm lam}}^2} \right)}
\end{equation}
in the asymptotic regime of turbulent burning \citep{Pocheau1994},
with $C_t = 4/3$. It is easy to observe that at $q = 0$, $v_{{\rm flame}} = v_{{\rm lam}}$
automatically and at $q >> v_{{\rm lam}}^2$, $v_{{\rm flame}} \rightarrow \sqrt{8q/3}$. 
The laminar flame speed is obtained by solving for the structure  
of a steady state deflagration wave. The method and numerical 
fit are presented in Appendix \ref{sec:defwave}.

\subsection{Neutrino Emission}

We use a open-source thermal neutrino 
emission subroutine\footnote{available in http://cococubed.asu.edu/code$\_$pageseos.shtml},
which calculates the neutrino luminosity $Q_{\nu}$ at a given temperature,
density, mean atomic number and mean proton number, using the 
analytic fit presented in \cite{Itoh1996}. Major thermal 
neutrino production channels are included, such as the 
pair-, photo-, plasma-, bremmstrahlung and recombination 
neutrino processes. 

To construct the neutrino spectra, we consider 
the neutrino emissivity due to different
neutrino generation mechanisms
$N_{{\rm neu}} = N_{{\rm pair}} + N_{{\rm plasma}} + ...$ . 
In our calculation, we focus on 
two major mechanisms: pair-annihilation 
process and plasma neutrinos.
The pair-annihilation neutrino 
emissivity at relativistic and non-degenerate 
regimes has an analytic fit \citep{Misiaszek2006} given by
\begin{equation}
N_{{\rm pair}} = F \phi(E_{\nu}),
\end{equation}
with 
\begin{equation}
F = \frac{G_F^2 m_e^8}{18 \pi^5} (M_-^{00} + M_+^{00})
\end{equation}
being the total number emissivity and
\begin{equation}
\phi( E_{\nu}) = \frac{A_1}{kT} \left( \frac{E_{\nu}}{kT} \right) {\rm exp}(-a E_{\nu} / kT)
\end{equation}
being the shape function. $E_{\nu}$ is the neutrino
energy. $m_e$ is the electron mass in MeV. $A_1$, $a$ and $\alpha$
are obtained from parametric fits of the exact relations. 
$G_F$ is the Fermi weak-coupling constant,
while $M_-^{00}$ ($M_+^{00}$) is the zeroth moment
of electron (positron) energy defined by 
\begin{eqnarray}
M_{\mp}^{nm} = (7 C_V^2 - 2 C_A^2) G^{\mp}_{n/2-1/2} G^{\mp}_{m/2-1/2} + \nonumber  \\
9 C_V^2 G^{\pm}_{n/2} G^{\mp}_{m/2} + (C_V^2 + C_A^2) \times \nonumber \\
(4 G^{\mp}_{n/2+1/2} G^{\pm}_{m/2+1/2} - G^{\mp}_{n/2-1/2} G^{\pm}_{m/2+1/2} - \nonumber \\
G^{\mp}_{n/2+1/2} G^{\pm}_{m/2-1/2}),
\end{eqnarray}
with
\begin{equation}
G^{\mp}_{n}(\alpha, \beta) = \frac{1}{\alpha^{3 + 2n}} \int_{\alpha}^{\infty} 
\frac{x^{2n + 1} \sqrt{x^2 - \alpha^2}}{1 + {\rm exp}(x \pm \beta)} dx
\end{equation}
being the Fermi integral and $\alpha$, $\beta$ and $x$
are the dimensionless electron mass, chemical potential 
and energy. $C_V$ and $C_A$ are the vector and axial
coupling constants.

There is also an analytic fit for plasma neutrinos 
given in \cite{Odrzywolek2007}
\begin{equation}
N_{{\rm plasma}}(E_{\nu}) = A_2 k T m_t^6 {\rm exp}(-E_{\nu} / k T),
\end{equation}
where 
\begin{equation}
A_2 = \frac{G_F^2 C_V^2}{8 \pi^4 \alpha h^4 c^9}
\end{equation}
and 
$m_t$ is the transverse electron mass related to 
the Fermi-Dirac distributions of electron $f_1$ and
positron $f_2$ by 
\begin{equation}
m_t^2 = \frac{4 \alpha}{\pi} \int_0^{\infty} \frac{p^2}{E} (f_1 + f_2) dp.
\end{equation}

In our simulations, the transverse electron mass
and the pair-annihilation total neutrino 
emissivity are tabulated
as functions of density and temperature. 
The emissivities of these two channels
can then be readily computed and summed
to obtain an energy distribution of neutrinos.

\section{Data Post-Processing}
\label{sec:post}

This section introduces the data processing 
after the hydrodynamics runs. They include the tracer
particle algorithm for reconstructing detailed 
nucleosynthesis yields, and a moment-integrated radiative 
transfer code for predicting the corresponding 
bolometric light curves.

\subsection{Particle Tracer}

The use of the 19-isotope network reaction products in 
tabular form only provides an approximation because of the absence
of electron capture and other 
off-$\alpha$-chain isotopes. In the literature, 
table forms of the energy production
and chemical composition are often used due to the 
significant difference between reaction and hydrodynamics 
time scales and length scales. However, such results 
are inadequate if we need to compare the nucleosynthesis with 
real observational data,
which can be recorded in fine details. 
In order to obtain a precise distribution of elements,
we use the tracer particle method presented 
in \cite{Travaglio2004} and \cite{Seitenzahl2010}. 
This method introduces a number of pseudo particles 
of either equal mass or equal volume. They follow but not affect
the fluid flow. 
The density and temperature of these particles are recorded
as functions of time. In simulations, the particle density,
temperature and velocity are obtained by bilinear interpolation
from the fluid properties. 

After the hydrodynamics simulations, the density-temperature
trajectory of each particle is recalled to trace back 
its chemical evolution. In our simulation, the reconstruction
is done based on the same 19-isotope network,
which can be compared with the results obtained
from hydrodynamics runs. 



\subsection{Radiative Transfer for Bolometric Light Curve}
\label{sec:method_radtran2}

After the simulations, we map the hydro results from
the two-dimensional Eulerian form to a
one-dimensional Lagrangian form with spherical
symmetry. During this transformation, the total mass, 
energy and momentum are conserved in order not to 
provide spurious perturbation to the profiles.
We follow the method in \cite{Zhang1994} to construct the light
curves by solving the one-dimensional time-dependent moment-integrated
radiative transfer equations, which include
two equations from Lagrangian hydrodyanmics,
\begin{eqnarray}
\frac{D v}{D t} = -\left( \frac{1}{\rho} \right) \frac{\partial (p + \nu_{{\rm ar}})}{\partial r} +
\frac{\chi_F}{c} F_r - \frac{G m(r)}{r^2}, \\
\frac{D \epsilon}{D t} + (p + \nu_{{\rm ar}}) \frac{D}{Dt} \left( \frac{1}{\rho} \right) = c \kappa_E E_r - 
4 \pi \kappa_P B(T) + \dot{Q}_{{\rm decay}}, 
\end{eqnarray}
and two equations from radiative transfer, 
\begin{eqnarray}
\frac{D}{Dt} \left( \frac{E_r}{\rho} \right) + 
\left[ f \frac{D}{Dt} \left( \frac{1}{\rho} \right) - (3f - 1) \frac{v}{\rho r} \right] E_r = \\
4 \pi \kappa_P B - c \kappa_E E_r - \frac{\partial (4 \pi r^2 F_r)}{\partial m}, \\
\frac{1}{c^2} \frac{D F_r}{Dt} + \frac{1}{q_{{\rm sph}}} 
\frac{\partial (f q E_r)}{\partial r} = - \frac{\chi_F \rho}{c} F_r
- \frac{2}{c^2} \left( \frac{v}{r} + \frac{\partial v}{\partial r} \right) F_r. 
\end{eqnarray}

The hydrodynamics part is almost identical to those in previous
parts. However, the physical quantities are defined 
in a staggered grid as typical in Lagrangian hydrodynamics.
Density $\rho$, specific internal energy $\epsilon$, fluid pressure $P$, 
enclosed mass $m(r)$, specific energy production rate 
due to the decay of radioactive isotopes $\dot{Q}_{{\rm decay}}$,
radiation-energy-mean opacity $\kappa_E$, Planck-mean opacity $\kappa_P$,
flux-mean opacity $\chi_F$, blackbody radiation rate $B$, Eddington factor 
$f$ and sphericity $q_{{\rm sph}}$ are defined on grid centers
($r = r_{n + 1/2}$, $n$ = 1, 2, 3...). 
Velocity $v$ and radiation flux $F_r$ are defined on grid boundaries
($r = r_n$, $n$ = 1, 2, 3 ...). 

$\nu_{{\rm ar}}$ is the artificial viscosity defined on the 
grid centers, which operates whenever fluid compression occurs.
At grid centers $k+1/2$, $\nu_{{\rm ar}}$ is proportional to the 
difference in the magnitudes of the velocities of nearest
grid boundaries, namely
\begin{equation}
\nu^n_{{\rm ar}~k+1/2} = 2 \nu_{{\rm num}}^2 (v^n_{k+1} - v^n_k)^2 \rho ^n_{k+1/2},
\end{equation}
whenever $D(1/\rho)/Dt < 0$ and $dv/dr < 0$
and $\nu_{{\rm num}} = 4$ is the numerical viscosity
coefficient, which controls the smearing of 
shocks.
The blackbody radiation rate is given by Planck's formula
$B(T) = (ac/4 \pi) T^4$. $\kappa_E$ and $\kappa_P$ 
are the radiation-energy-mean opacity and 
Planck-mean opacity, given by 
\begin{eqnarray}
\kappa_E = \int \kappa(\nu) E_{\nu} d\nu / \int E_{\nu} d\nu, \\
\kappa_P = \int \kappa(\nu) B_{\nu}(T) d\nu,
\end{eqnarray}
with $B_{\nu}(T) = 2 h \nu^3 / (e^x - 1)$ 
being the Planck distribution function
at some given temperature $T$, frequency $\nu$
and $x = h \nu / k T$. 
$\chi_F$ is the radiation-flux-mean opacity, given by 
\begin{equation}
\chi_F = \int \chi(\nu) F_{\nu} d\nu / F_r. 
\end{equation}
Since we consider only bolometric luminosity,
which means all physics quantities are 
frequency-integrated, we follow the choice of 
\cite{Hoeflich1993} that $\kappa_E = \kappa_P$,
and $\chi_F$ is replaced by the Rosseland 
mean opacity $\kappa_R$, defined by
\begin{equation}
\kappa_R = \frac{\int \frac{1}{\chi_{\nu}} \frac{\partial B}{\partial T} d\nu}{\int \frac{\partial B}{\partial T} d\nu}.
\end{equation}
We describe how we model the 
continuum opacity source and include
the atomic transition lines as
opacity sources in Appendix \ref{sec:opa}.

The Eddington factor is an algebraic closure of the radiative
transfer equations, given by 
\begin{eqnarray}
f_{\nu} = \left( \int I_{\nu} \mu^2 d\mu \right) / \left( \int I_{\nu} d\mu \right),
\label{eq:edd_fac}
\end{eqnarray}
with $\mu = {\rm cos} \theta$ being the direction cosine of radiations
with respect to the radial direction. $I_{\nu}$ is the monochromatic 
radiation intensity at frequency $\nu$. The sphericity $q_{{\rm sph}}(\nu)$
measures the level of isotropy of light rays, defined by
\begin{equation}
{\rm ln} (q_{\nu}) = \int^r_{r_c} \left[ (3 f_{\nu} - 1) / (r' f_{\nu}) dr' \right],
\end{equation}
with $r_c$ being the radius, within which the matter
is optically thick.

In our simulations, $\dot{Q}_{{\rm decay}}$ is the energy release
due to decays of $^{56}$Ni and $^{56}$Co. In the radiative
transfer, we consider only the time evolution of three
isotopes: $^{56}$Ni, $^{56}$Co and $^{56}$Fe. 
The energy release is related to the decay rates given by
\begin{equation}
\dot{Q}_{{\rm decay}} = D_{\gamma} \left( \epsilon^{\gamma}_{{\rm Ni}} \frac{DX_{{\rm Ni}}}{Dt} 
+ \epsilon^{\gamma}_{{\rm Co}} \frac{DX_{{\rm Co}}}{Dt} \right) + 
\epsilon^{e^+}_{{\rm Co}} \frac{DX_{{\rm Co}}}{Dt},
\end{equation}
where $\epsilon^{\gamma}_{{\rm Ni}}$ and $\epsilon^{\gamma}_{{\rm Co}}$
are the specific energy release due to the decays of 
the isotopes by emitting gamma rays, and $\epsilon^{e^+}_{{\rm Co}}$
is that by emitting positrons. The amount of isotopes
can be computed analytically, 
\begin{eqnarray}
X_{{\rm Ni}} = X_{{\rm Ni (ini)}} \exp(-t/\tau_{{\rm Ni}}), \\
X_{{\rm Co}} = X_{{\rm Ni (ini)}} \frac{\tau_{{\rm Co}}}{\tau_{{\rm Co}} - \tau_{{\rm Ni}}} 
\left[ \exp(-t/\tau_{{\rm Co}}) - \exp(-t/\tau_{{\rm Ni}}) \right],
\end{eqnarray}
with $\tau_{{\rm Ni}} = 7.605 \times 10^5$ s
and $\tau_{{\rm Co}} = 9.822 \times 10^6$ s
the decay half-lives of the two radioactive isotopes. 
$X_{{\rm Ni}}$ and $X_{{\rm Co}}$ are the mass
fractions of nickel-56 and cobalt-56. 
The energy released by the positron decay channel is 
assumed to be completely absorbed by local matter 
because a positron has a much higher scattering 
cross section than a photon, even at low density.
Only a fraction of photons is absorbed, which is controlled by the 
deposition function $D_{\gamma}$. To determine the 
deposition function, we follow the prescription described in
\cite{Swartz1995} by solving analytically the grey
radiative transfer of the gamma ray radiations in the 
two-stream approximations. Specifically, we obtain
the energy-integrated intensity for both incoming $(I^-)$
and outgoing directions $(I^+)$ by solving 
\begin{equation}
\pm \frac{\partial I^{\pm}_{\gamma}}{\partial z} = \eta - \kappa_{\gamma} \rho I^{\pm},
\end{equation}
subject to boundary conditions $I^- = 0$ at the surface
of the ejecta and $I^- = I^+$ at the core. $\eta$ is the 
frequency integrated gamma ray emissivity. 
After solving 
for $I^{\pm}$ at each grid shell, the deposition function
is computed by 
\begin{equation}
D_{\gamma} = \frac{4 \pi \kappa_{\gamma} J}{\dot{Q}_{{\rm decay}}},
\end{equation}
with $J$ being the moment-integrated intensity
\begin{equation}
J = \frac{1}{4 \pi} \oint I d\Omega,
\end{equation} 
which is integrated over all solid angles $d\Omega$.

In Eq. (\ref{eq:edd_fac}), we need to solve 
$I_{\nu}$ formally.
We decompose it into the symmetric part 
$j_{\nu} = [I_{\nu}(r,\mu) + I_{\nu}(r,-\mu)] / 2$
and anti-symmetric part $h_{\nu} = [I_{\nu}(r,\mu) - I_{\nu}(r,-\mu)] / 2$. 
They satisfy
\begin{eqnarray}
\frac{1}{c} \frac{D j_{\nu}}{Dt} + \frac{\partial h_{\nu}}{\partial s} = 
\kappa_{\nu} \rho B_{\nu}(T) - \notag \\
 \left[ \kappa_{\nu} \rho + (1 - 3 \mu^2) \frac{v}{r c} - 
\frac{(1 + \mu^2)}{c} \frac{D {\rm ln} \rho}{Dt} \right] j_{\nu},  
\label{eq:j_rad} \\
\frac{1}{c} \frac{D h_{\nu}}{Dt} + \frac{\partial j_{\nu}}{\partial s} = 
-\chi_{\nu} \rho h_{\nu} - \frac{2}{c} \left( \frac{v}{r} + 
\frac{\partial v}{\partial r} \right) h_{\nu}.
\label{eq:h_rad}
\end{eqnarray}
In the above equations, $s^2 = r^2 - p^2$ where 
$p$ is the impact factor of the ray. 
By solving Eqs. (\ref{eq:j_rad}) and (\ref{eq:h_rad})
with boundary conditions $j_{\nu}(r = R) = h_{\nu}(r = R)$
and $h_{\nu} (r = 0) = 0$ for all $s$, we may solve 
for $I_{\nu}(r,\nu)$, which then 
provides the Eddington factor $f_{\nu}$ and sphericity $q_{\nu}$. 

To solve the radiative transfer equations, we input the 
density, velocity, temperature and composition profiles.
Since the explosion does not yet achieve the homologous
expansion, we map the initial velocity profile to a
homologously expanding one with the same kinetic energy
by using the results from the end of the simulations
of the explosion phase. Certainly, a consistent 
way to obtain a homologous profile
is to let the system evolve. This requires either a
sufficiently large simulation box or box size that 
varies with time. However, the first way requires 
an impractically large amount of computational
resource, while most of it is not used except at
later time when the star starts to expand. 
On the other hand, with the second way we
can keep track of the 
fluid motion while using a manageable computer
resource, but it requires special numerical
treatment of physics components that are sensitive
to the resolution, including the sub-grid turbulence
and the level-set method. Therefore, at this stage, 
we artificially replace the velocity profile by 
a homologous one which conserves the total energy of 
the system.




\section{Hydrodynamics Results}
\label{sec:results}

This section presents the hydrodynamics results 
of various code tests. They include one-dimensional and
two-dimensional code tests, such as the 
standard shocktube tests, Gresho vortex 
and tests of hydrodyanmics instabilities. 
They aim at testing the validity of the code and
its capability in shock-capturing, level of
accuracy and numerical diffusion. 
Then we present hydrodynamics from explosion 
models including the PTD and DDT models. 
Though it is known that the PTD model cannot 
provide a healthy explosion that matches 
typical SNIa ejecta, its slowness in flame propagation
allows various hydrodynamics instabilities
to form, which can be used to check the physical components
collectively, such as the level-set method, 
sub-grid turbulence and products of the
deflagration wave. The DDT 
model is one of the possible SNIa mechanisms.
Also, our results can be compared directly for models 
with similar configurations in the literature. 

\subsection{One-dimensional Code Test}

\begin{table*}
\begin{center}
\caption{Input parameters for one-dimensional code tests.}
\begin{tabular}{|c|c|c|c|c||c|c|c|c|}
\hline
Test & $t_0$ & $x_0$ & $\rho_L$ & $v_L$ & $p_L$ & $\rho_R$ & $v_R$ & $p_R$ \\ \hline
1 & 0.2 & 0.3 & 1.0 & 0.75 & 1.0 & 0.125 & 0.0 & 0.1 \\
2 & 0.5 & 0.5 & 1.0 & -2.0 & 0.4 & 1.0 & 2.0 & 0.4 \\
3 & 0.035 & 0.4 & 5.99924 & 19.5975& 460.894 & 5.99242 & -6.19633 & 46.0950 \\
4 & 0.012 & 0.8 & 1.0 & -19.59745 & 1000.0 & 1.0 & -19.59745 & 0.01 \\ \hline

\end{tabular}
\label{table:Test1D}
\end{center}
\end{table*}

In this two-dimensional code, we study the 
one-dimensional limit by reducing the number
of grids in either one dimension to a few grids
and assuming Cartesian coordinates, i.e. the 
volume of each grid is independent of the radial
distance from the axis of symmetry. 
We follow the tests provided in \citep{Toro1997}
which are all shocktube tests to understand
whether the code can correctly evolve various types of 
shocks. They all have exact solutions
computed by Riemann solvers.
In this part, we use a polytropic EOS with the ratio
of specific heat $\gamma = 1.4$. The EOS is written 
as $p = \rho \epsilon (\gamma - 1)$. The simulation
takes place in a spatial domain of interval $[0,1]$
with 2000 computing cells. The Courant number is fixed
to be 0.5. The initial data consists of two parts, the 
left-hand state $(\rho_L, v_L, p_L)$ and the 
right-hand state $(\rho_R, v_R, p_R)$. 
The transition takes places at $x_0$ and the 
simulation is run for a duration $t_0$. We tabulate the 
input configurations in Table \ref{table:Test1D}.

\begin{figure}
\centering
\includegraphics*[width=8cm, height=6cm]{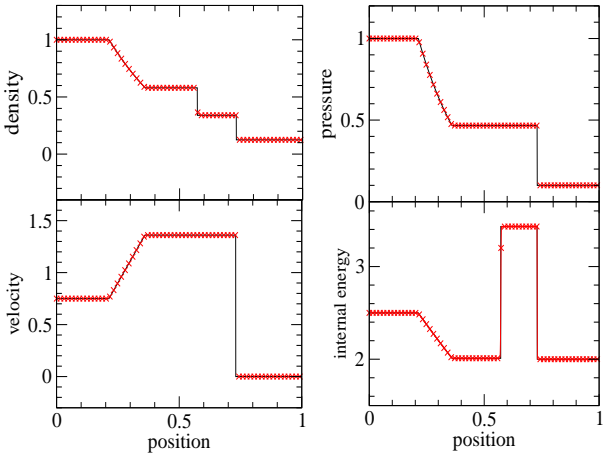}
\caption{Density, pressure, velocity and internal energy
in Test 1 at $t_0 = 0.2$ and $x_0 = 0.3$.}
\label{fig:codetest1D_test1}
\end{figure}

\begin{figure}
\centering
\includegraphics*[width=8cm, height=6cm]{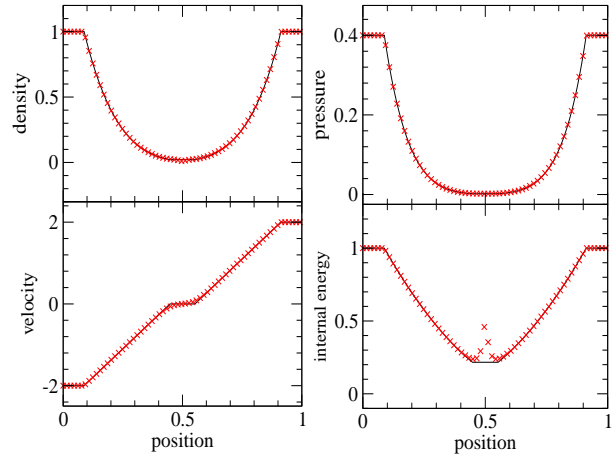}
\caption{Density, pressure, velocity and internal energy
in Test 2 at $t_0 = 0.5$ and $x_0 = 0.5$.}
\label{fig:codetest1D_test2}
\end{figure}

In Fig. \ref{fig:codetest1D_test1}, we plot the 
density, pressure, velocity and internal energy of the fluid
for Test 1 at $t = 0.2$. This test is a modified version of 
Sod's test which assesses the entropy satisfaction property
of the numerical scheme. The final solution includes
a right shock wave, a right-traveling wave and a left sonic 
rarefaction wave. As seen from the figure, the code can 
very well preserve the sharp edge of the 
shock and contact discontinuity. Also, the entropy glitch, 
which usually appears in low-order schemes
in a bump inside the rarefaction, does not appear. 

\begin{figure}
\centering
\includegraphics*[width=8cm, height=6cm]{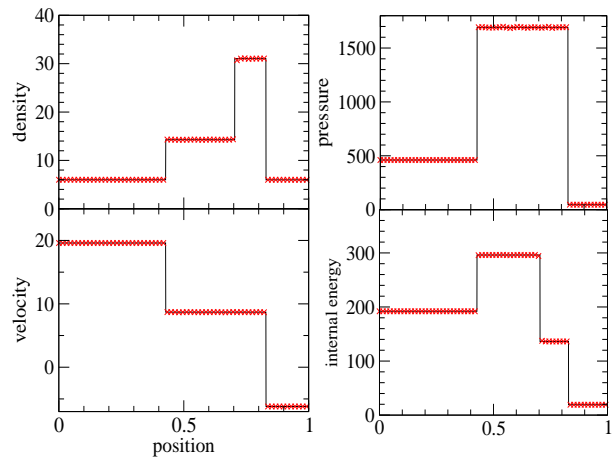}
\caption{Density, pressure, velocity and internal energy
in Test 3 at $t_0 = 0.035$ and $x_0 = 0.4$.}
\label{fig:codetest1D_test3}
\end{figure}

\begin{figure}
\centering
\includegraphics*[width=8cm, height=6cm]{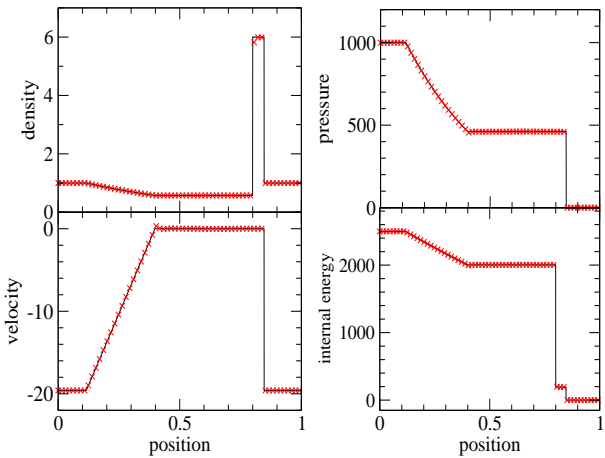}
\caption{Density, pressure, velocity and internal energy
in Test 4 at $t_0 = 0.012$ and $x_0 = 0.8$.}
\label{fig:codetest1D_test4}
\end{figure}

In Fig. \ref{fig:codetest1D_test2}, we plot
similarly the density, pressure, velocity and internal 
energy of the fluid for Test 2 at $t_0 = 0.15$. This test contains 
a solution of two rarefaction waves and a stationary
contact wave. The solution contains also a low-density region,
and it tests how well the code can handle 
low-density fluid flow. The numerical solution 
follows the analytic solution very well, except for an
undershoot in the velocity near $x = 0.5$ and 
an overestimation by a factor of 2
in the internal energy. Despite these, the performance 
of the WENO scheme is in general much better than 
other low-order schemes, which result in spurious oscillations
in the velocity field and a larger overestimation in the
internal energy (See \cite{Toro1997} for a detailed
comparison among different numerical schemes). 

In Fig. \ref{fig:codetest1D_test3} we plot the test results 
for Test 3 
at $t_0 = 0.035$. This test assesses the 
robustness of a code in processing strong discontinuities.  
There are two right-traveling shock waves, a right-traveling 
contact discontinuity, and a left-traveling shock. 
It can be seen from the figure that 
all discontinuities can be resolved without smearing. 

In Fig. \ref{fig:codetest1D_test4} we plot the density,
pressure, velocity and internal energy in Test 4
at $t_0 = 0.012$. This test
challenges the code in evolving a slowly-moving or 
stationary shock discontinuity. The final solution contains
a left-traveling rarefaction wave, right-traveling 
shock wave and a stationary contact discontinuity. 
The numerical solution largely follows the analytic solution
except for a small overshoot in pressure, velocity
and internal energy at $x = 0.4$ at the edge of rarefaction wave. 
Also, there are spurious oscillations with a small amplitude  in 
density at $x = 0.8$.

Combining these four tests, we can see that our code performs well 
in treating numerical discontinuity and it can
mimic closely the analytic solution. Typical numerical 
inaccuracies, including the entropy glitches, spurious oscillations
and smearing are highly suppressed.

\subsection{Two-dimensional Code Test}

The one-dimensional tests aim at validating the 
shock-capturing properties of the WENO scheme 
under different situations. In this part, we 
extend the code test to two dimensions. 
Besides shock-capturing, we also study
the numerical diffuseness and the advection of smooth 
flow of the code. 
In all tests the polytropic EOS identical to the 
one-dimensional code tests is used with specific
heat ratio $\gamma = 1.4$. The Courant 
number is taken to be 0.5.

\begin{table*}
\begin{center}
\caption{Fractional changes in mass $\Delta M/M$ 
and energy $\Delta E/E$ in $Gresho$ test under 
different resolutions. $L^1 (p - p_{{\rm ref}})$
is the $L^1$ norm of the error in the fluid pressure, 
with reference to the initial state.}
\begin{tabular}{|c|c|c|c|c|c|c|c|}
\hline
$\Delta$ & $N_x$ & $N_y$ & $\Delta M / M$ & $\Delta E / E$ & $L^1 (p - p_{{\rm ref}})$ \\ \hline
$2 \times 10^{-2}$ & 50 & 50 & $2.96 \times 10^{-3}$ & $4.08 \times 10^{-4}$ & $4.50 \times 10^{-3}$ \\
$1 \times 10^{-2}$& 100 & 100 & $1.09 \times 10^{-3}$ & $1.52 \times 10^{-4}$ & $1.63 \times 10^{-3}$ \\
$5 \times 10^{-3}$ & 200 & 200 & $1.49 \times 10^{-5}$ & $2.10 \times 10^{-5}$ & $2.89 \times 10^{-4}$ \\
$2.5 \times 10^{-3}$ & 400 & 400 & $1.94 \times 10^{-6}$ & $2.73 \times 10^{-6}$ & $3.26 \times 10^{-5}$ \\ \hline
\end{tabular}
\label{table:Test2D_test1}
\end{center}
\end{table*}

\begin{figure}
\centering
\includegraphics*[width=9cm, height=7cm]{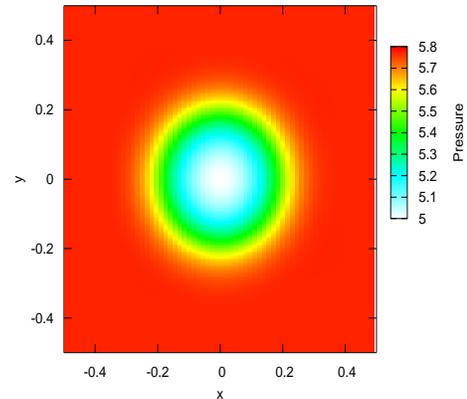}
\caption{Density contour plot of $Gresho$ test at $t = 3$.}
\label{fig:codetest2D_test1}
\end{figure}

The first test ${Gresho}$ \citep{Gresho1990} is to 
evaluate the code performance in handling 
time-independent solution, which tests the code
in handling advection of smooth functions. The simulation takes place in a 
square box of $-0.5<x<0.5$ and $-0.5<y<0.5$ with $\Delta$ = 0.02, 
0.01, 0.005 and 0.0025 units. The boundary is free everywhere. 
The simulation time is 3 (code unit).
The initial profile is a steady solution
given by $\rho = 1$ everywhere,
$v(r) = 5r$, $p(r) = 5 + 25/2 r^2$
at $0 < r <0.2$, $v(r) = 2 - 5r$, $p(r) = 9 + 4 {\rm ln}(r/0.2) 
+ 25/2 r^2 - 20r$ at $0.2 < r< 0.4$ and $v(r) = 0$ and 
$p(r) = 3 + 4 {\rm ln}2$ at $0.4 < r < 0.5$. Here $r$ is the 
distance from the origin. Obviously, any departure of the final
solution from the initial one must be due to numerical 
errors from the advection and numerical diffusion. 
In Fig. \ref{fig:codetest2D_test1} we plot
the linear density contour plot at $t = 3$ for $dt = 0.005$. 
It can be seen that the circular structure is maintained. 
The outer part remains in uniform pressure. In Table 
\ref{table:Test2D_test1} we list the numerical errors in
total mass, energy and the $L^1$-norm of the fluid pressure.
At high resolutions, the results show a third order 
convergence which is consistent with the order of 
the time-discretization scheme of the code.

\begin{figure}
\centering
\includegraphics*[width=9cm, height=7cm]{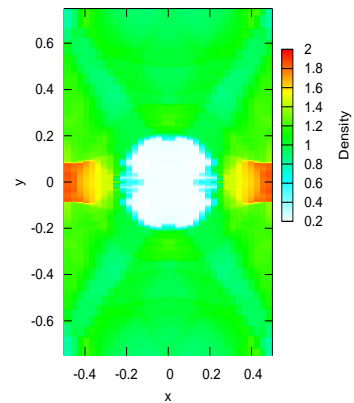}
\caption{Density contour plot of $Blast$ test at $t = 1$.}
\label{fig:codetest2D_test2}
\end{figure}

\begin{figure}
\centering
\includegraphics*[width=9cm, height=7cm]{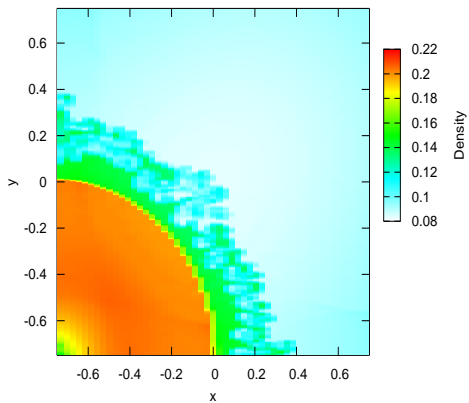}
\caption{Pressure contour plot of $Explosion$ test at $t = 2.667$.}
\label{fig:codetest2D_test3}
\end{figure}

\begin{figure}
\centering
\includegraphics*[width=9cm, height=7cm]{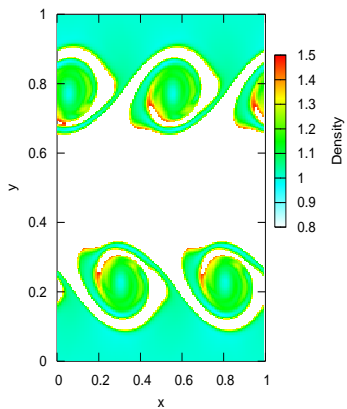}
\caption{Density contour plot of KH test at $t = 1$.}
\label{fig:codetest2D_test4}
\end{figure}

The second test is the {\it Blast} test \citep{Toro1997}
which takes place in a
rectangular box of dimensions $-0.5<x<0.5$ and $-0.75 < y < 0.75$
with reflecting boundaries everywhere. 
The resolution is $512 \times 768$. The initial condition
is given by $p = 10$ at $r < 0.1$ and $p = 1$ at $r > 0.1$. 
$\rho = 1$ everywhere. This test studies again the 
capability of the code in handling the low density region,
the sharpness of the hydrodynamical instabilities and 
the shock-shock, or shock-contact discontinuity 
interactions. 
In Fig. \ref{fig:codetest2D_test2} we plot the linear
density contour at $t = 1.0$. It can be seen 
that the symmetry along the x-axis and along 
the y-axis are preserved. Also, the Richter-Meshkov instability
at the boundary of contact-discontinuity 
in the form of dense filaments or fingers are 
sharply captured.

The third test is the $Explosion$ test \citep{Toro1997} 
which is very similar to SNIa explosion 
but without nucleosynthesis. The simulation 
box is $512 \times 512$ with dimensions $0 < x < 1.5$ and 
$0 < y < 1.5$. The inner boundaries are reflecting and the 
outer boundaries are free. The initial condition is 
given as an overpressure region at the core, with 
$p(r) = 1.0$ and $\rho(r) = 1.0$ at $r < 0.4$ and 
$p(r) = 0.1$ and $\rho(r) = 0.125$ at $r > 0.4$. 
In Fig. \ref{fig:codetest2D_test3} we plot the 
linear density contours at $t = 2.667$. 
At this moment, the weak reflected shock has arrived 
inside the shock sphere, and the unstable contact 
discontinuity of the surface in the form of 
plumes can be observed. Again, a sharp symmetry 
with respect to a reflection about 
$x = y$ can be observed. 

The last test is the Kelvin-Helmholtz (KH) 
test \citep{McNally2011} which studies the 
KH instability. The simulation box is 
$256 \times 256$ with dimensions $0 < x < 1$ and 
$0 < y < 1$ with periodic boundaries everywhere. 
The simulation is done up to $t = 2$. The initial condition
is that two fluids of different densities
flow in opposite directions with 
a perturbation: $p = 1$ everywhere, $\rho = 2 $
at $0.25 < y < 0.75$ and $\rho = 1$ otherwise.
An initial velocity is given as
$v_y = A (1 + {\rm cos}(2 \pi x)) (1 + {\rm cos}(2 \pi y))$.
In Fig. \ref{fig:codetest2D_test4} we plot the linear
density contours at $t = 1$. The KH instability
in the form of spiral shape outside the denser region 
can be observed. 

Combining these four tests, we have seen that 
the code performs 
satisfactorily with the 
desired accuracy and shock-capturing ability in 
multi-dimensional runs. Also, the code has
shown its ability in handling hydrodynamical
instabilities, such as the Kelvin-Helmholtz instability. 

\subsection{Code Test for PTD models}
\label{sec:results_PTD}

In this part, we return our focus to 
code test of SNIa explosions using the PTD
models. The simulation uses the Helmholtz 
EOS \citep{Timmes1999a} and all the 
simulations are done in cylindrical 
coordinates with 500 grids in each 
direction. Each grid has a size 
$\Delta = dr = dz \approx 11$ km. A hydrostatic
equilibrium model with a given density is 
constructed with an isothermal profile
of $T = 10^8$ K and a composition of 
$X(^{12}$C$) = X(^{16}$O$) = 0.5$. 
The code enforces
a minimum density of $10^5$ g cm$^{-3}$,
which is also the atmospheric density.  
The time step is limited by the
Courant-Friedrich-Lewy condition:
\begin{equation}
\Delta t_{{\rm max}} = c_{cfl} \Delta / (c_s + |v_r| + |v_z|),
\end{equation}
with $c_{cfl} = 0.5$ and $c_s$ the local 
sound speed. 

An initial three-finger flame front is imposed,
which is comparable with the c3 flame reported in
\cite{Niemeyer1996} and \cite{Reinecke1999b}.
The reason we choose this flame front is the
same as that in the literature: we want to bypass
the slow laminar flame stage and consider the 
stage where Rayleigh-Taylor instabilities are important. 
The matter enclosed by the flame front is 
first burnt. We follow the choice in \cite{Niemeyer1995b}
to set the initial specific turbulence kinetic energy 
$q = 10^{10}$ cm$^{2}$ s$^{-2}$. We also set this value 
to be the minimum $q$ so as to avoid unphysical result
when calculating the production and dissipation of $q$.

\begin{table*}
\begin{center}
\caption{Simulation setup for the test of PTD models. 
Lengths are in units of 
km and densities are in units of 
$10^{9}$ g cm$^{-3}$. Energy is in units
of $10^{50}$ erg and masses are in units
of solar mass.}

\begin{tabular}{|c|c|c|c|c|c|c|c|}
\hline
Model & $dx$ & $\rho_c$ & mass & radius & $E_{{\rm nuc}}$ & $E_{{\rm kin}}$ & $M_{{\rm Ni}}$ \\ \hline
2D-1a-PTD & 11.0 & 3.0 & 1.377 & 1475.0 & 7.141 & 2.109 & 0.320 \\ \hline
\label{table:TestPTD}
\end{tabular}
\end{center}
\end{table*}

\begin{figure}
\centering
\includegraphics*[width=8cm, height=6cm]{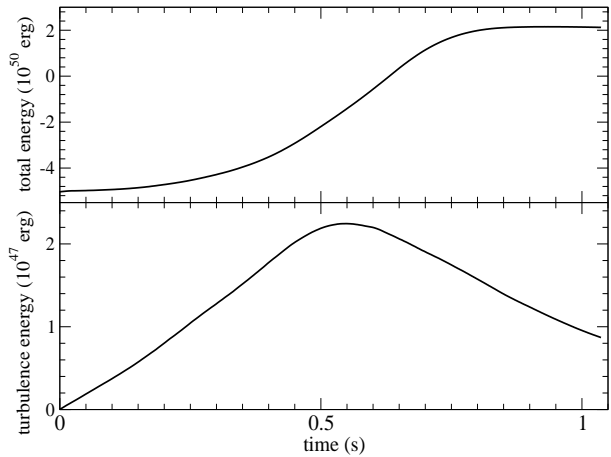}
\caption{Upper panel: Total energy against time for Model 2D-1a-PTD.
Lower panel: Total turbulence kinetic energy against time for the same model.}
\label{fig:model1_2d_PTD_energy}
\end{figure}

\begin{figure}
\centering
\includegraphics*[width=8cm, height=6cm]{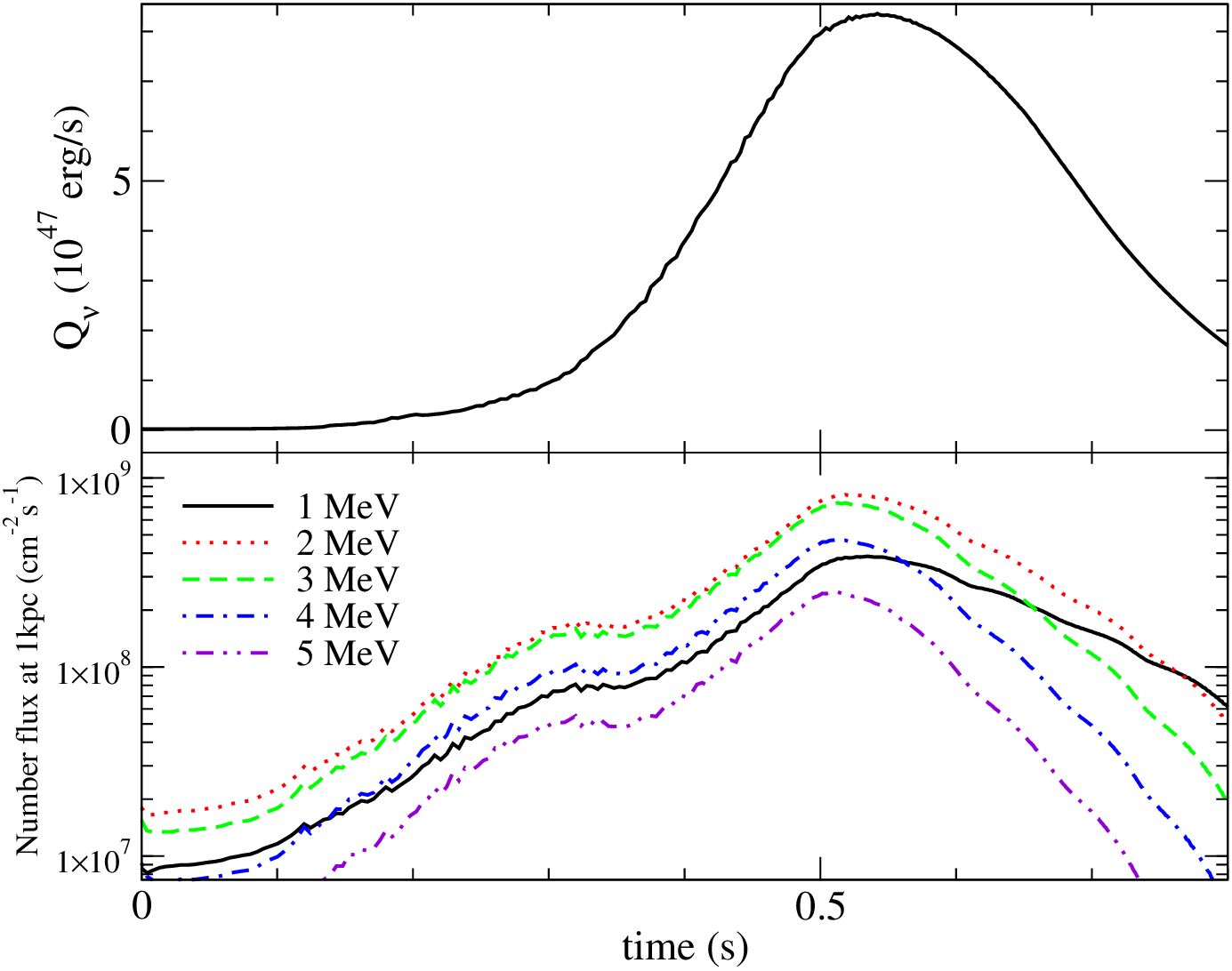}
\caption{Upper panel: Total neutrino luminosity against time
for Model 2D-1D-PTD. Lower panel: Neutrino energy spectra 
against time for the same model.}
\label{fig:model1_2d_PTD_energyloss}
\end{figure}

We tabulate the 
simulation setups in Table. \ref{table:TestPTD}.
We plot in the upper panel of
Fig. \ref{fig:model1_2d_PTD_energy} the total energy
against time for Models 2D-1a-PTD. At early time 
the energy release relies on slow laminar flame, thus
the energy growth is slow. At about $t \approx 0.5$ s,
the sub-grid turbulence has developed and it boosts the 
propagation of flame and enhances the consumption of 
fuel significantly. At about 0.7 s, the nuclear
flame has released enough energy to balance the 
gravitational energy. At $t \approx 0.8$ s,  the
expansion of the star leads to density decrease at 
the flame front, which lowers the energy output of the flame
as well. So, the total energy levels off and reaches 
a constant at about $2 \times 10^{50}$ erg. Notice that our 
initial model and related physics are chosen to mimic 
those of the Model $c3-2d-256$ in \cite{Reinecke2002a}. 
Our results are comparable with theirs, 
which also give the final energy at $2 \times 10^{50}$ erg. 
In the lower panel, we plot 
the total turbulence kinetic energy $q$ against time.  
Similar to the neutrino luminosity, the sub-grid 
turbulence energy reaches its maximum at around 0.6 s. 
We plot in the upper panel of 
Fig. \ref{fig:model1_2d_PTD_energyloss} the 
neutrino luminosity against time. 
The model shows a single peak at about 0.5 s. 
This implies that the burning is the most vigorous
at that moment. In the lower panel we plot the 
neutrino energy spectra with energy 1 MeV - 5 MeV as
functions of time. In early time, the neutrino 
energy peaks at 2 MeV. At later time, when the WD
starts to expand, the 1 MeV neutrino flux
becomes the dominant one. 

\begin{figure}
\centering
\includegraphics*[width=8cm, height=6cm]{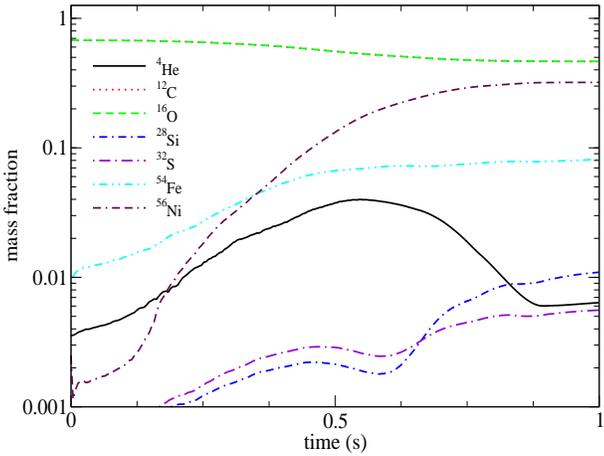}
\caption{Isotope mass fraction against time for Model 2D-1a-PTD.}
\label{fig:model1_2d_PTD_xiso}
\end{figure}

\begin{figure}
\centering
\includegraphics*[width=9cm, height=7cm]{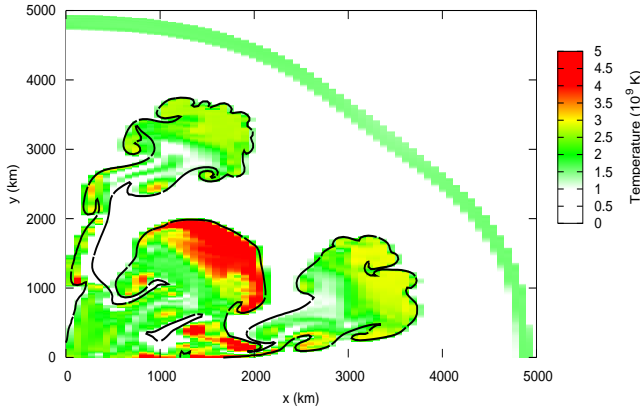}
\caption{The deflagration and detonation fronts at $t = 1$ second
for Model 2D-1a-PTD.}
\label{fig:model1a_2d_ptd_flame8}
\end{figure}

We also plot in Fig. \ref{fig:model1_2d_PTD_xiso}
the time evolution of chemical isotopes. 
At early time, iron-peaked elements are produced,
and a considerable amount of $^{4}$He is produced.
At later time, when the fluid element expands
and its temperature drops, $^4$He recombines to $^{56}$Ni
again, and incomplete burning causes
the production of a trace amount of IME. The results can 
also be compared with Model $c3-2d-256$ \citep{Reinecke2002a} 
which observes a total of 0.109 $M_{\odot}$ IME
and 0.40 $M_{\odot}$ iron-peaked elements. 
We obtain similar amount of iron-peaked elements
but the IME abundance is lower than theirs. 
This may be due to the difference in approximating
the ash composition and the energy release at 
lower density, which is the main site for 
the production. 
We plot in Fig. \ref{fig:model1a_2d_ptd_flame8}
the flame shape at $t = 1$ s. The initial
flame is the $c3$ front shape as described
in \cite{Niemeyer1995b}, which is 
a "three fingers" shape. The flame shows
signatures of hydrodynamical 
instabilities: the Rayleigh Taylor instability in the form
of mushroom shape at the top
of the "fingers" and KH instability in the 
form of curly shape along the "fingers". Injection
of fuel into the flame can also be seen between
"fingers".

\subsection{Code test for DDT models}
\label{sec:results_DDT}

\newpage

\newpage

In this part we present results for the DDT
model. The DDT model is actually a natural 
continuation of the PTD model for two 
reasons. First, before the detonation starts,
the evolution is exactly the same as the PTD 
model. Second, the criterion for the first detonation spot
relies on the local turbulence that
the flame width $\delta$ equals the Gibson
length, also known as the turbulence length
scale $l_{{\rm Gib}}$, below which the laminar flame is 
the dominant propagation channel.
Mathematically, we have 
\begin{equation}
\delta = l_{{\rm Gib}} = \Delta (v_{{\rm lam}} / 2 q)^{3/2}. 
\end{equation}
The flame width is obtained from the results 
by solving the deflagration wave, as described
in Appendix \ref{sec:defwave}, which is 
a function of density. We note that this
relation, which is equivalent to the Karlovitz number
$Ka = 1$, only states that turbulence starts to 
destroy the flame front. But as argued in 
\cite{Niemeyer1997b}, this mechanism can allow the 
heat from the ash to be transported to the fuel
much more efficiently than only by conduction. 
This creates a much wider region with
sufficient temperature for carrying out 
carbon burning simultaneously, which is believed to be 
one of the keys for triggering the detonation.
In the simulation, whenever there is a grid
on the flame front satisfying this criterion, a
spherical detonation spot of radius $\Delta$
is artificially placed and evolved similarly
as the deflagration front. However, because 
the deflagration front has burnt partially
the stellar material in the core, whenever
the detonation front reaches the flame surface,
we assume that the detonation front stops. 

\begin{table*}
\begin{center}
\caption{Simulation setup for the test of DDT models. 
Lengths are in units of 
km and densities are in units of 
$10^{9}$ g cm$^{-3}$. Energy is in units
of $10^{50}$ erg and masses are in units
of solar mass. $t_{{\rm tran}}$ is the 
first DDT time in seconds. }

\begin{tabular}{|c|c|c|c|c|c|c|c|c|}
\hline
Model & $dx$ & $\rho_c$ & mass & radius & $E_{{\rm nuc}}$ & $E_{{\rm kin}}$ & $M_{{\rm Ni}}$ & $t_{{\rm tran}}$ \\ \hline
2D-1a-DDT & 11.0 & 3.0 & 1.377 & 1475.0 & 15.651 & 10.618 & 0.733 & 0.70 \\ \hline
\label{table:TestDDT}
\end{tabular}
\end{center}
\end{table*}

\begin{figure}
\centering
\includegraphics*[width=8cm, height=6cm]{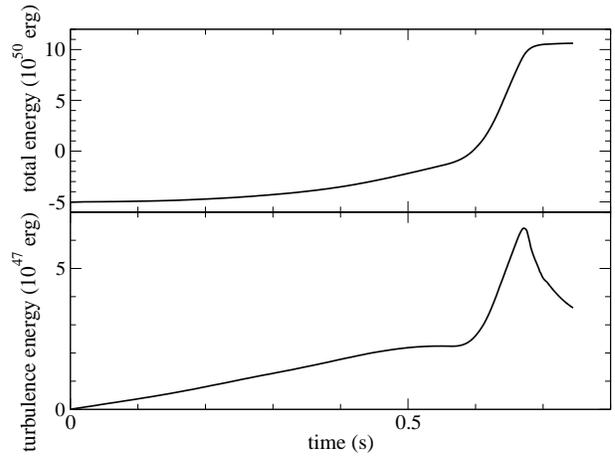}
\caption{Upper panel: Total energy against time for Model 2D-1a-DDT.
Lower panel: Total turbulence kinetic energy against time
for the same model.}
\label{fig:model1_2d_DDT_energy}
\end{figure}

\begin{figure}
\centering
\includegraphics*[width=8cm, height=6cm]{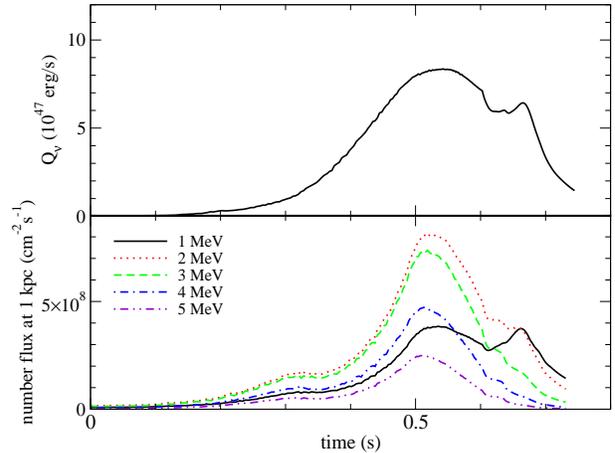}
\caption{Upper panel: Total neutrino luminosity against time
for Model 2D-1a-DDT. Lower panel: Neutrino energy spectra 
against time for the same model.}
\label{fig:model1_2d_DDT_energyloss}
\end{figure}

We carry out tests for models using DDT and we tabulate
the input parameters and final explosion energetics in 
Table \ref{table:TestDDT}. The explosion energetics
can be compared with Model $Z3a$ in \cite{Golombek2005}. But 
the major difference between Model $Z3a$  
and ours is that the former one allows the 
detonation front to pass through burnt
regions, which is not allowed in ours in view of some
direct numerical simulations of the instability of 
the shock front in the presence of ash product \citep{Maier2006}. 
They reported $E_{{\rm nuc}} = 14.4 \times 10^{50}$ erg with 
$E_{{\rm kin}} = 8.70 \times 10^{50}$ erg. 
The DDT transition time is 0.85 s 
and the final amount of $^{56}$Ni in their 
model is 0.65 $M_{\odot}$. Our model shows a higher 
energy release, a higher $^{56}$Ni production
and an earlier trigger in DDT.

\begin{figure}
\centering
\includegraphics*[width=8cm, height=6cm]{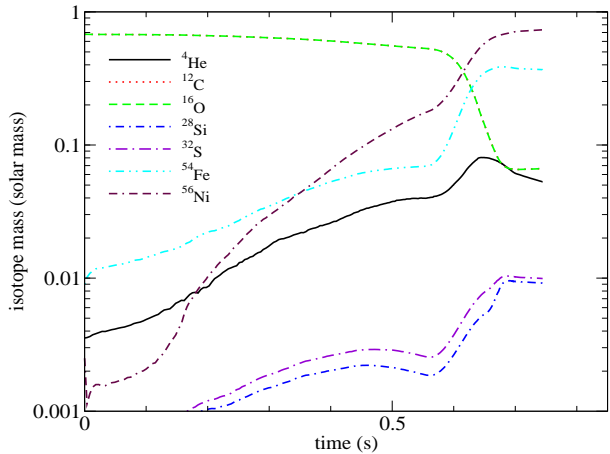}
\caption{Isotope mass fraction against time for model 2D-1a-DDT.}
\label{fig:model1_2d_DDT_xiso}
\end{figure}

\begin{figure}
\centering
\includegraphics*[width=8cm, height=6cm]{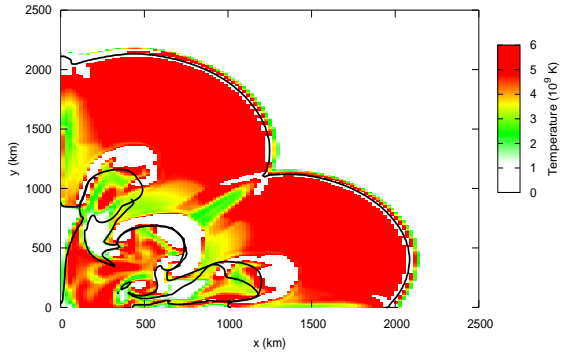}
\caption{The deflagration and detonation fronts of Model 2D-1a-DDT at 
$t = 1.00$. The temperature distribution is represented
in different colours. }
\label{fig:model1a_2d_DDT_flame8}
\end{figure}

In the upper panel of Fig. 
\ref{fig:model1_2d_DDT_energy}, we plot the total
energy against time for Model 2D-1a-DDT.
Prior to the detonation front being triggered, 
the evolution is identical to that of the PTD model. 
By comparing with Model 2D-1a-PTD, we see that our model is
consistent with models in the literature that most energy
is released by the detonation front 
instead of the deflagration front,
which is also needed to match the 
ejecta velocity with the observed SNIa. 
In the lower panel, we plot 
the total turbulence kinetic energy. 
The turbulence constantly develops as the fluid 
motion becomes chaotic, and then it drops slightly
when the WD expands. Turbulence 
grows again when the detonation is launched,
which reaches its maximum as the explosion
has burnt all the fuel in the WD. But it then drops
sharply due to the rapid expansion of the star.

In the upper panel of Fig. \ref{fig:model1_2d_DDT_energyloss} we plot
the neutrino luminosity of Model 2D-1a-DDT.
The neutrino luminosity is decreasing when  
the detonation is just started. But it goes
up again due to a large amount of matter being
thermalized. In the lower panel we plot the 
neutrino energy spectra with energy 1 MeV - 5 MeV
as functions of time. Again the number flux
of each energy band shows a rise after
DDT has started and after $t = 0.7$ s, 
the 1 MeV neutrino number flux exceeds the 
other four fluxes of higher energy bands. 
We also plot the isotope abundances against time
in Fig. \ref{fig:model1_2d_DDT_xiso}. The early
time evolution is comparable to the PTD model. 
After the detonation has started, much more 
iron-peaked elements and IME are produced, leaving
trace amounts of $^{12}$C and $^{16}$O. 
At last, we plot deflagration and detonation
flame front at $t = 1$ s in Fig. \ref{fig:model1a_2d_DDT_flame8}. 
The flame front is shaped by the fluid flow,
that "mushroom caps" and injection of 
fuel can be observed from the flame front.
The detonation front
first starts at the outermost part of the 
flame front, which develops at first spherically,
but then loses its symmetry when the front
encounters another detonation front and 
the deflagration front.

\section{Observational Prediction}
\label{sec:obs}

We have demonstrated in previous sections that our
hydrodynamics code can provide explosion 
energetics comparable with models in the literature. 
To further study the hydrodynamics result, we 
perform a series of post-processing procedures to 
extract SNIa observables, including
the nucleosythesis yields by the tracer particle method and
the bolometric light curve using a moment-integrated 
radiative transfer code. 

\subsection{Nucleosynthesis Products}

We use the tracer particle method to keep track of 
the density and temperature of each fluid 
parcel, which represents a fixed amount  
of mass elements in the fluid. We repeat the 
2D-1a-DDT model but with different numbers of 
tracer particles. 
We study its numerical convergence by adjusting the 
initial radial and angular separation of the fluid
parcels. In the first test, we use a simple 19-isotope
network with a fixed output time of hydrodynamics
properties at 0.025 s. We 
tabulate the nucleosynthesis yields in Table 
\ref{table:Run2D_tracer_results}. We also 
include the chemical composition from
the hydrodynamics run for comparison. In general,
the post-processed yields are different from that 
of the hydrodynamics run in two ways; the final
distributions of $^{12}$C and $^{16}$O are 
not the same in the post-processed results.
This is related to the slow-burning in the 
low density regime, which is not assumed 
in the deflagration wave product. This also 
affects the IME distribution so that much higher 
mass fractions of $^{28}$Si and $^{32}$S 
are observed. Second, the amount of iron-peaked
elements are drastically different in that 
the $^{54}$Fe are generally lower, showing that
lower amount of matter has reached nuclear
statistical equilibrium for a
complete burning comparing to the expectation from
the deflagration wave solution. 
We also compare how the number of tracers
along each direction affects the final results. 
We find that for $N = 80^2$ the changes in the
final yields differ by less than 1 $\%$, while more
abundant isotopes such as $^{54}$Fe and $^{56}$Ni
generally converge faster than less 
abundant elements such as $^{40}$Ca and $^{44}$Ti.

\begin{table*}
\begin{center}
\caption{Masses of all isotopes at the end of simulations
for different numbers of tracer particles
$N = 40^2$, $80^2$ and $160^2$. The approximate
yields from the hydrodyanmics run are also 
included for comparison. The 19-isotope
network is used. The record time
is fixed at  $t = 0.025$ s.
All quantities are in units of solar mass.}

\begin{tabular}{|c|c|c|c|c|}
\hline
Isotope & Hydro & $N = 40^2$ & $N = 80^2$ & 
		  $N = 160^2$ \\ \hline
$^{12}$C & $6.68 \times 10^{-2}$ & $1.61 \times 10^{-2}$ & $1.61 \times 10^{-2}$ & 
		   $1.60 \times 10^{-2}$ \\
$^{16}$O & $6.68 \times 10^{-2}$ & $3.74 \times 10^{-2}$ & $4.36 \times 10^{-2}$ & 
		   $4.33 \times 10^{-2}$ \\
$^{24}$Mg & $1.60 \times 10^{-5}$ & $8.36 \times 10^{-3}$ & $1.11 \times 10^{-2}$ & 
			$1.06 \times 10^{-2}$ \\
$^{28}$Si & $9.17 \times 10^{-3}$ & $1.97 \times 10^{-1}$ & $1.99 \times 10^{-1}$ & 
			$1.98 \times 10^{-1}$ \\ 
$^{32}$S & $9.93 \times 10^{-3}$ & $9.22 \times 10^{-2}$ & $9.16 \times 10^{-2}$ &  
		   $9.08 \times 10^{-2}$ \\
$^{36}$Ar & $6.90 \times 10^{-3}$ & $1.93 \times 10^{-2}$ & $1.90 \times 10^{-2}$ & 
			$1.88 \times 10^{-2}$ \\
$^{40}$Ca & $1.27 \times 10^{-2}$ & $1.82 \times 10^{-2}$ & $1.78 \times 10^{-2}$ & 
			$1.76 \times 10^{-2}$ \\
$^{44}$Ti & $3.80 \times 10^{-4}$ & $9.96 \times 10^{-5}$ & $9.73 \times 10^{-5}$ & 
			$9.66 \times 10^{-5}$ \\
$^{48}$Cr & $2.91 \times 10^{-2}$ & $8.71 \times 10^{-4}$ & $8.46 \times 10^{-4}$ & 
			$8.39 \times 10^{-4}$ \\
$^{52}$Fe & $4.20 \times 10^{-2}$ & $2.10 \times 10^{-2}$ & $2.07 \times 10^{-2}$ & 
			$2.06 \times 10^{-2}$ \\
$^{54}$Fe & 0.368 & 0.100 & $9.89 \times 10^{-2}$ & $9.99 \times 10^{-2}$ \\
$^{56}$Ni & 0.733 & 0.836 & 0.831 & 0.831 \\ \hline
\label{table:Run2D_tracer_results}
\end{tabular}
\end{center}
\end{table*}


\subsection{Bolometric Light Curve}

\begin{figure}
\centering
\includegraphics*[width=8cm, height=6cm]{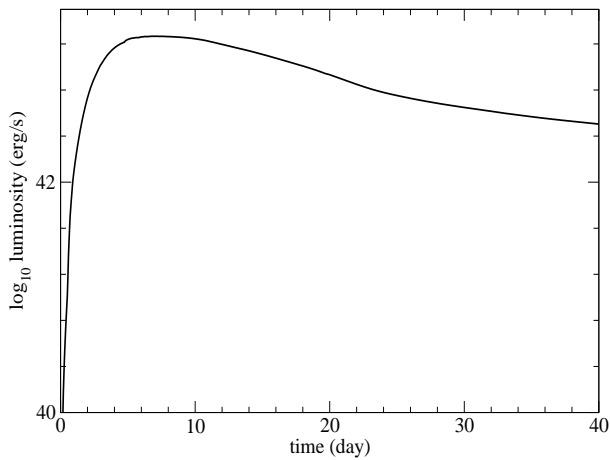}
\caption{Bolometric light curves of
Model 2D-1a-DDT.}
\label{fig:lc_model1_ddt}
\end{figure}

Using the hydrodynamics results 
from models as we have obtained
in Sections \ref{sec:results_PTD} and 
\ref{sec:results_DDT}, we can predict the 
expected observational signals of these explosions.
We plot in Fig. \ref{fig:lc_model1_ddt} 
the bolometric light curves for Models
2D-1a-DDT using the 
one-dimensional radiative transfer code.
The nickel abundance and its distribution
have stronger effects on the peak luminosity and 
width. We obtain a peak
luminosity log$_{10}$ $L_{{\rm peak}} = 43.1$ with 
a drop of the absolute magnitude at 15 days after
maximum $\Delta m_{15} = 0.58$. A mild but observable
secondary bump can be seen at about Day 30, showing that
the photosphere has receded inside the iron 
layer. Beyond Day 40, when $^{56}$Co decay becomes the dominant 
energy source, the curve drops at a constant rate.  

\section{Discussion and Conclusion}
\label{sec:discussion}

In this paper, we have presented a two-dimensional 
Eulerian hydrodynamics code for modeling SNIa using 
cylindrical coordinates with the level-set method 
as the flame capturing scheme. We also included the sub-grid 
turbulence for modeling turbulent flame
and a neutrino subroutine 
for computing the corresponding
neutrino spectra. We also used a tracer particle
scheme for nucleosynthesis and developed
a moment-integrated radiative transfer 
code for computing the bolometric light curve.
We have preformed benchmark tests in one
and two dimensions to evaluate the accuracy 
and robustness of the code. Also, we have 
studied the SNIa explosion using the 
PTD and DDT models. The nucleosynthesis details 
and the bolometric light curves of our 
simulation models are comparable to those 
similar explosion models presented in the literature. 

However, it should be noted that 
a consistent comparison with models
in other works are difficult for two reasons. 
First, there are different numerical 
treatments for the same physical process.
For example, in earlier works
the effective flame propagation speed is chosen to be
the maximum of various flame propagation  
mechanisms, while in later works the 
transition of the flame from laminar to turbulent 
is based on analytic models \citep{Pocheau1994}. 
However, how the flame propagation varies with
local turbulence is not yet exactly known. 
Second, each code has its own prescription for 
the input physics, such as the EOS and
the nuclear reaction network, while some processes, such as the 
nucleosynthesis and the turbulence model, 
depend on these models. For example, 
we use the same EOS presented in \cite{Timmes1999a}
as the FLASH code \citep{Calder2002},
but their model does not include the effects
of sub-grid turbulence. Also, our nuclear 
energy production depends on the choice of 
EOS. We directly use a nuclear reaction network to 
calculate the energy released by the 
deflagration and detonation waves, while in 
\cite{Schmidt2005} all materials are assumed to 
be burnt into nuclear statistical 
equilibrium at high densities and 
only up to $^{24}$Mg at low densities. 
The difference has a subtle effect on the explosion
since the produced energy may boost the fluid motion
for the creation of sub-grid velocity fluctuations
and at the same time makes the star expand. 
The produced energy then affects the local density
which consequently affects the energy release again.
Therefore it is difficult to carry out a quantitative
comparison when the hydrodynamics, flame physics and 
nucleosynthesis are coupled. 

In future work, we plan to include other 
numerical components, such as the 
advection-diffusion-reaction scheme 
and other models of sub-grid turbulence.
By comparing the results of these numerical
components we may better understand their roles
in the explosion, as well as the limitation 
and validity regimes of
these models.  
Also, as a natural continuation of our previous 
work \citep{Leung2013}, the effects of dark matter
will be included into the dynamics. We plan to 
examine whether the presence of dark matter
affects the role of SNIa as a standard candle,
which is critical in the discovery of dark energy 
and measurements of cosmological distances.

\section{Acknowledgment}
\label{sec:ack}

We thank K.-W. Wong for his development 
of a 1D WENO solver based on which our 2D SNIa 
code is constructed. We also thank F. X. Timmes for 
his open-source codes of the Helmholtz EOS, nuclear 
reaction network and neutrino emission. This work is 
partially supported by a grant from the Research Grant
Council of the Hong Kong Special Administrative Region, 
China (Project No. 400910). SCL is supported by the 
Hong Kong Government through the Hong Kong PhD Fellowship.

\appendix
\section{Deflagration wave}
\label{sec:defwave}

In Sections \ref{sec:methods_LSM} and 
\ref{sec:Network} we mentioned that deflagration
and detonation are coupled to the hydrodynamics
as energy sources. Here we describe the 
governing equations for deflagration,
how they are solved to obtain the results.

Deflagration is one of the combustion processes where
electron scattering is the major heat transport mechanism. 
The process is so slow that the fluid remains almost
isobaric. The wave structure and the flame speed are 
studied in \cite{Timmes1992a}. At least four 
methods can be used to study the
structure of deflagration wave. They include
direct simulations, diffusion approximations, 
eigenvalue method and variational approximations,
arranged in descending order of information
provided. Direct simulation is the computationally most
expensive method, but is the most accurate one,
because no assumption is made in the evolution 
equations. On the other hand, 
variational approximation is the least accurate as 
it gives only a lower bound of the flame speed 
and reveals no information about the flame structure.  
In our simulations, we use the diffusion approximations in finding the 
flame structure and its corresponding speed,
which is shown to be a good approximation due
to the slow propagation of flame compared
with the much faster sound speed. 

The diffusion approximation starts from the 
general Euler equations in spherical coordinate
and Lagrangian formulation, namely,
\begin{eqnarray}
\frac{\partial m}{\partial r} &=& 4 \pi r^2 \rho, \\
\frac{Dv}{Dt} &=& - 4 \pi r^2 \frac{\partial P}{\partial m} - \frac{G M}{r^2}, 
\label{eq:defwave_vel}\\
\frac{D\epsilon}{Dt} + P \frac{\partial (1/\rho)}{\partial t} &=& \frac{1}{\rho} \frac{\partial}{\partial r} 
\left( \chi \frac{\partial T}{\partial r} \right) + \dot{q}, 
\label{eq:defwave_eps} \\
\dot{q} &=& N_{{\rm A}} \sum_i \frac{DY_i}{Dt} B_i, \\
\frac{DY_i}{Dt} &=& \sum_{j,k} -Y_i Y_j \lambda_{jk} (i) + Y_l Y_k \lambda_{kj} (l). 
\end{eqnarray} 
$B_i$ and $Y_i$ are the binding energy and 
the number fraction of isotope $i$. $\lambda_{jk}$
is the reaction rate of nuclear reaction from 
isotope $j$ to $k$. Other variables have the
the same physical meaning as in the main text. The derivative 
$D/Dt$ is the time derivative in the frame moving
with the fluid. The diffusion approximation
then assumes that the fluid has always the same 
pressure. Gravity is neglected because
the typical size of deflagration reaction
zone is much smaller than the density scale height.
This means that the velocity equation Eq.
(\ref{eq:defwave_vel}) can be neglected.

In the heat diffusion term on the right hand side 
of Eq. (\ref{eq:defwave_eps}) $\chi$ is the total thermal conductivity,
which consists of contributions from 
both electron and photon conductivities
\begin{equation}
\chi = \chi_e + \chi_{\gamma}.
\end{equation}
Photon conductivity can be exactly found 
as $\chi_{\gamma} = 4 a c T^3 / 3 \kappa \rho$,
with $a = 7.5657 \times 10^{-15}$ erg cm$^{-1}$ K$^{-4}$
being the radiation constant, $c$ being the speed of light
and $\kappa$ being the opacity.

To find the electron conductivity, we follow the 
prescription of \cite{Khokhlov1997}. First we have $\chi_e$
expressed in terms of the effective electron collisional frequency $\nu_e$
\begin{equation}
\chi_e = 4.09 \times 10^9 T \frac{x^3}{\sqrt{1 + x^2}} \left( \frac{10^{16}}{\nu_e} \right) {\rm ergs~cm^{-1}s^{-1}K^{-1}},
\end{equation}
with $x = p_F/m_e c$ being the dimensionless  
Fermi momentum. The electron collisional frequency
can be separated into ion-electron and electron-electron 
collisional frequencies by 
\begin{equation}
\nu_e = \nu_{ee} + \nu_{ei}.
\end{equation}

The electron-ion collisional frequency is given by 
\begin{equation}
\nu_{ei} = 1.78 \times 10^{16} \sqrt{1 + x^2} \frac{\bar{Z} \Lambda}{\bar{A} Y_e},
\end{equation}
with $\bar{A}$ being the mean atomic mass and 
$\bar{Z}$ being the mean atomic number. $Y_e$ is the 
electron fraction and $\Lambda$ is the Coulomb logarithm. 
\begin{eqnarray}
\Lambda &=& \ln \left[ 
\left( \frac{2 \pi Z}{3} \right)^{1/3} \sqrt{\frac{3 \Gamma + 3}{2}} \right] -
\frac{x^2}{2 (1 + x^2)} + \nonumber \\
&& \frac{\pi}{2} \alpha \beta^2 
\frac{1 + 1.3 \alpha}{1 + \alpha^2 (0.71 - 0.54 \beta^2)}.
\end{eqnarray}
Here, $\Gamma = 2.275 \times 10^5 \bar{Z}^{5/3} (\rho Y_e)^{1/3} / T$, 
$\beta = x / \sqrt{1 + x^2}$ and $\alpha = Z / 137 \beta$. 

\begin{figure}
\centering
\includegraphics*[width=8cm, height=6cm]{figA1.eps}
\caption{Isotope abundances against
density for 13 isotopes. See the 
text for the lists of elements.}
\label{fig:abundance_13iso}
\end{figure}

\begin{figure}
\centering
\includegraphics*[width=8cm, height=6cm]{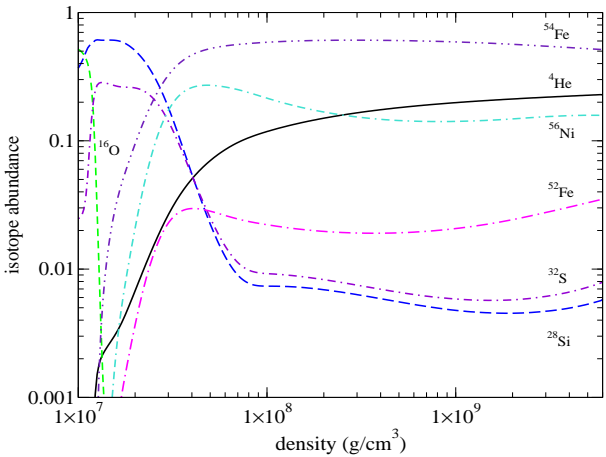}
\caption{Isotope abundances against
density for 19 isotopes. See the 
text for the lists of elements.}
\label{fig:abundance_19iso}
\end{figure}

Similarly, we have the electron-electron collision frequency 
as a function of temperature and density
\begin{equation}
\nu_{ee} = 0.511 T^2 \frac{x^{3/2}}{(1 + x^2)^{5/4}} J(y)~{\rm s^{-1}},
\end{equation}
with $y = \sqrt{3} T_{pe}/T$. $T_{pe}$ is the electron-plasma 
temperature given by
\begin{equation}
T_{pe} = 3.307 \times 10^8 \frac{x^{3/2}}{(1 + x^2)^{1/4}}~{\rm K}.
\end{equation}
$J(y)$ is an integral that cannot be evaluated analytically and is given
numerically by
\begin{equation}
J(y) = \frac{1}{3} \left( \frac{y}{1 + ay} \right)^3 \ln \left( \frac{2 + by}{y} \right),
\end{equation}
where $a= 0.113$ and $b = 1.247$. 

Using these information, we find numerically the 
deflagration wave propagation speed, the energy production
and the ash composition. To start the deflagration wave,
we follow the prescription in \cite{Timmes1992a}. We first
ignite the innermost grid cells, setting its initial temperature
sufficient for nuclear runaway. Then we start the 
evolution and gradually a steady deflagration wave forms. 
The laminar flame velocity can be obtained by studying the 
motion of the steady wave structure. The ash composition
is obtained from the regions swept by the deflagration wave.
The internal energy production is calculated by comparing the initial
and final specific internal energy.

\begin{figure}
\centering
\includegraphics*[width=8cm, height=6cm]{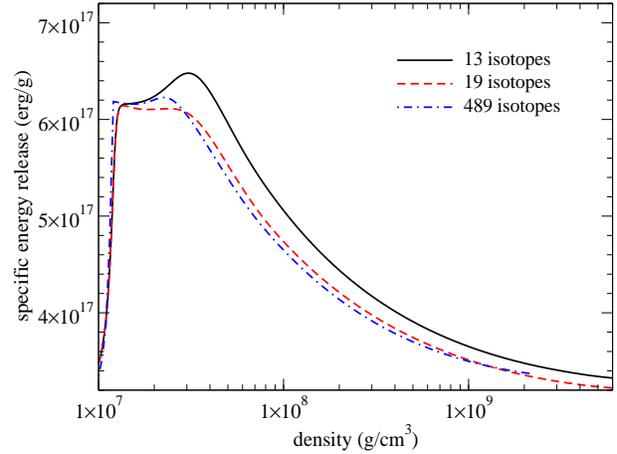}
\caption{Specific internal energy production against
density for 13, 19 and 489 isotopes. See the 
text for the lists of elements.}
\label{fig:energy_19iso}
\end{figure}

As shown in \cite{Timmes1992a}, the laminar flame
velocity can be well approximated by
\begin{equation}
v_{{\rm lam}} = 92.0~{\rm km~s}^{-1} 
\left( \frac{\rho}{10^9~{\rm g~cm}^{-3}} \right)^{0.805} \left( \frac{X_C}{0.5} \right)^{0.5},
\end{equation}
with $X_C$ being the mass fraction of $^{12}$C.  

We plot in Figs. \ref{fig:abundance_13iso}
and \ref{fig:abundance_19iso} the ash composition
against fuel density using 13-isotope and 19-isotope 
networks. At low density  $\rho \sim 10^7$ g cm$^{-3}$, 
only carbon is consumed. Oxygen remains barely unchanged. 
Magnesium and silicon are the major products. 
At $\rho \sim 5 \times 10^7$ g cm$^{-3}$, oxygen is also 
consumed, with silicon and sulphur being the major products. 
At density around $10^8$ g cm$^{-3}$, iron and nickel become
prominent. Also helium is produced due to photo-disintegration. 
$^{54}$Fe is the major product 
when it is included in the nuclear reaction network, which is 
absent in the 13-isotope network. 

In Fig. \ref{fig:energy_19iso} we plot the energy
release against fuel density using 13-, 19- and
489-isotope networks. The specific
internal energy production (SIEP)
is in the order of $10^{17}$ erg. At low density $\rho \sim 10^7$ g cm$^{-3}$,
SIEP increases with density and the oxygen mass fraction decreases.
This is because the reaction rate increases with density. 
At density above $5 \times 10^7$ g cm$^{-3}$,
SIEP drops. This does not mean less fuel is consumed. 
As mentioned in \cite{Reinecke2002a}, at high temperature,
the highly endothermic photo-disintegration $^{56}$Ni $\rightarrow 14^{4}$He
becomes important, which compensates for the rise of SIEP.

\section{Detonation Wave}
\label{sec:detwave}

To determine the wave structure, we follow
the prescription in \cite{Sharpe1999}.
We first find the thermodynamics state of 
shocked fluid from a given state. Then, 
using the shocked state as an initial condition, 
the wave structure can be obtained by solving
the steady state limit of Euler equations
with nuclear reactions.

For the first part, the upstream and downstream 
states are related by the Rankine-Hugoniot conditions
\begin{gather}
\rho u = \rho_0 D, \\
p + \rho v^2 = p_0 + \rho_0 D^2, \\
\epsilon + \frac{p}{\rho} + \frac{v^2}{2} = \epsilon_0 + \frac{p_0}{\rho_0} + \frac{D^2}{2}.
\end{gather}
Quantities with a subscript 0 are the upstream (unshocked) state.
$D$ is the outflow velocity. Both $\epsilon_0(\rho_0, T_0, X_0)$
and $p(\rho_0, T_0, X_0)$ are dependent on given density, temperature
and composition. In particular, $\rho_0$ and $T_0$ are 
free parameters, but the actual results are insensitive to $T_0$,
because the upstream electrons are highly degenerate. 
$X_0$ is the composition of fuel, namely 
0.5 $X_C$ and 0.5 $X_O$. In general, the final composition $X$
is a set of quantities (depending on the 
choice of isotopes) to be determined. 

After having the post-shock state, we  
determine the reaction zone size and the ash composition
by assuming a one-dimensional planar and steady flow, namely
\begin{gather}
v \frac{d \rho}{dx} + \rho \frac{du}{dx} = 0, \\
\rho v \frac{du}{dx} + \frac{dp}{dx} = 0, \\
\frac{d \epsilon}{dx} - \frac{p}{\rho^2} \frac{d \rho}{dx} = 0, \\
\frac{dX}{dx} = \frac{R}{v},
\end{gather}
where $R$ is the reaction rate. Then, using 
\begin{equation}
dp = \left( \frac{\partial p}{\partial \rho} \right)_{T,X} d \rho + 
\left( \frac{\partial p}{\partial T} \right)_{\rho,X} dT +
\sum^N_{i=1} \left( \frac{\partial p}{\partial X_i} \right)_{\rho,T} dX_i,
\end{equation}
and 
\begin{equation}
d \epsilon = \left( \frac{\partial \epsilon}{\partial \rho} \right)_{T,X} d \rho + 
\left( \frac{\partial \epsilon}{\partial T} \right)_{\rho,X} dT +
\sum^N_{i=1} \left( \frac{\partial \epsilon}{\partial X_i} \right)_{\rho,T} dX_i,
\end{equation}
the Euler equations with nuclear reactions
in the steady state limit can be reduced to 
\begin{eqnarray}
\frac{d \rho}{dx} &=& -\frac{\rho a^2_f}{v} 
\frac{{\bf \sigma} \cdot {\bf R}}{\eta}, \\
\frac{dT}{dx} &=& \left( \frac{\partial p}{\partial T} \right)^{-1}_{\rho, X} 
\left\lbrace \left[ u^2 - 
\left( \frac{\partial p}{\partial \rho} \right)_{T,X} \right]  
\frac{d \rho}{dx} - \right. \nonumber \\
&& \hspace{1cm} \left. \sum^N_{i=1} \left( \frac{\partial p}{\partial X_i} 
\right)_{\rho,T,X_{j \neq i}} \frac{dX_i}{dx} \right\rbrace , \\
\frac{dY}{dx} &=& \frac{R}{v},
\label{eq:deton_struct}
\end{eqnarray}
where
\begin{equation}
\eta = a^2_f - v^2
\end{equation}
is the sonic parameter, 
\begin{equation}
a^2_f = \left( \frac{\partial p}{\partial \rho} \right)_{T,X} + 
\left[ \frac{p}{\rho^2} - \left( \frac{\partial \epsilon}{\partial \rho} \right)_{T,X} \right]
\left( \frac{\partial p}{\partial T} \right)_{\rho,X} \left( \frac{\partial \epsilon}{\partial T} \right)^{-1}_{\rho,X}
\end{equation}
is the sound speed of constant composition (also known
as frozen sound speed) and
\begin{eqnarray}
\sigma_i &=& \frac{1}{\rho a^2_f} \left\lbrace \left( \frac{\partial p}{\partial X_i} \right)_{\rho,T,X_{j \neq i}} - 
\left( \frac{\partial p}{\partial T} \right)_{\rho, X} 
\left( \frac{\partial \epsilon}{\partial T} \right)^{-1}_{\rho,X} \times \right.
\nonumber \\
&& \left. \left[ \left( \frac{\partial \epsilon}{\partial X_i} 
\right)_{\rho,T,X_{j \neq, i}} - \left( 
\frac{\partial q}{\partial X_i} \right)_{X_{j \neq i}} \right] \right\rbrace
\end{eqnarray} 
is the thermicity constant, such that 
${\bf \sigma} \cdot {\bf R}$ is the thermicity. 

Eq. (\ref{eq:deton_struct}) can be integrated 
into the reaction zone. Notice that there 
are actually two types of detonation structure
implied from these equations. One type is that
throughout the detonation wave both $\eta$ and 
thermicity are positive, while the other type is that
both quantities are both positive or negative
simultaneously. The first type is the 
well-known Chapman-Jouguet (CJ) detonation, while 
the second type is called the supported pathological
(SP) detonation.

CJ detonation occurs at low density ($\rho < 2 \times 10^7$ g cm$^{-3}$).
The boundary condition is given by
$\eta = {\bf \sigma} \cdot R = 0$ 
at $x \rightarrow \infty$. This corresponds to the 
ash propagating at frozen sound speed at the 
point where no more nuclear reaction can proceed. 

For SP detonation ($\rho > 2 \times 10^7$ g cm$^{-3}$), the integration is divided into
two sections. First, we integrate the equation 
close to $\eta \rightarrow 0$ (while thermicity
needs not to be zero at that position). Assume
$\eta = 0$ at $x = x_p$ and we have integrated up to
$x = x_p - \Delta x_p$, by making use of the 
symmetry of Eqs. (\ref{eq:deton_struct}) from the 
zero-velocity point and the continuity of 
thermodynamics variables, we obtain the post-zero-point
states at $x = x_p + \Delta x_p$, where
\begin{equation}
\rho(x_p + \Delta x_p) = \rho(x_p - \Delta x_p),
\end{equation}
and
\begin{eqnarray}
T(x_p + \Delta x_p) &=& T(x_p - \Delta x_p) - \nonumber \\
&& \sum^N_{i=1} \left( \frac{\partial p}{\partial X_i} 
\right)_{\rho,T,X_{j \neq i}} \frac{dX_i}{dx} (2 \Delta x_p).
\end{eqnarray}
After the above transition, the integration can be 
carried on again until no net nuclear reaction continues. 
To determine the correct eigenvalue, we require 
$\eta = {\bf \sigma} \cdot R = 0$ at the same position.

\section{Flame Capturing using the Point-Set method}
\label{sec:methods_PSM}

In two-dimensional simulations, the flame front is 
represented by a system of line segments. 
The point-set method (see for example in 
\citep{Glimm1981, Glimm1985, Glimm1987, Glimm1988, Glimm2002} for applications
in two-dimensional systems, \citep{Glimm1999, Glimm2000, Glimm2003}
for applications in three-dimensional systems and
\citep{Zhang2009} for its application in SNIa modeling.)
introduces a set of pseudo-particles, which form a line by assuming that 
the nearest neighbors are linked. Similar to the level-set
method, the particles are transported by the fluid advection 
with a speed $\vec{v}_{{\rm fluid}}$
and its own propagation with velocity $v_{{\rm flame}} \hat{n}$ , such that
\begin{equation}
\vec{v}_{{\rm node}~i} = v_{{\rm flame}} \hat{n}_i + \vec{v}_{{\rm fluid}}.
\end{equation}
$\hat{n}_i$ is the unit normal vector of the 
flame front pointing from the $i$-th node towards the fuel.
The fluid velocity is 
given directly by the Euler equations. The 
flame propagation depends on the fluid density 
and the local normal direction of flame front 
pointing towards the fuel. 
The normal of a node is defined by the positions
of its closest neighboring nodes. For example, 
given a node $i$ with position $(x_i, y_i)$, with a
distance to its next node $i+1$ given by 
\begin{equation}
d_{i,i+1} = \sqrt{(x_i - x_{i+1})^2 + (y_i - y_{i+1})^2},
\end{equation}
we define the normal direction of the line 
segment formed by the nodes $i$ and $i+1$ by
\begin{equation}
\hat{n}_i = \left( -\frac{y_{i+1} - y_i}{d_{i,i+1}}, 
\frac{x_{i+1} - x_i}{d_{i,i+1}}\right).
\end{equation}
Notice that whether the normal vector pointing
towards the fuel or ash is arbitrary. In our study,
we choose the former one. 

After all node velocities are found, the 
positions are updated by 
\begin{equation}
\vec{x}_{{\rm node~(new)}} = \vec{x}_{{\rm node~(old)}} +
\vec{v}_{{\rm node}} \Delta t_{{\rm node}}.
\end{equation}
There is no accelerating term for the 
node because the particles serve only as 
markers of the deflagration front for computing 
the change in energy and isotopes, which do not
interact with the fluid directly. 

The time step may not be the same as that
of the hydrodynamics one, because the resolution of 
the point-set method and the hydrodynamics are independent. 
In general, we require 
\begin{equation}
\Delta t_{{\rm node}} = C \frac{l_{{\rm min}}}{v_{{\rm node}}},
\end{equation}
with $C$ a positive number smaller than unity
and $l_{{\rm min}}$ the 
minimum distance between neighboring nodes. 

Apart from $l_{{\rm min}}$, there are two more parameters
controlling the point-set resolution, 
$l_{{\rm max}}$ and $l_{{\rm merge}}$. They 
define the inter-node maximum separation and 
merging distance between non-neighboring nodes. 
These parameters are essential in maintaining a
consistent resolution of the surface during 
the evolution, because the distance between 
linked nodes can become too far or close, or there
can be lines crossing each other. These 
phenomena occur frequently when the fluid motion is 
turbulent. In those cases, node addition, 
removal or reconnection is needed. Notice that
in the literature, node reconnection is also 
known as surface splitting/merging in the 
context of multi-dimensional simulations.

\begin{figure}
\centering
\includegraphics[width=8cm, height=6cm]{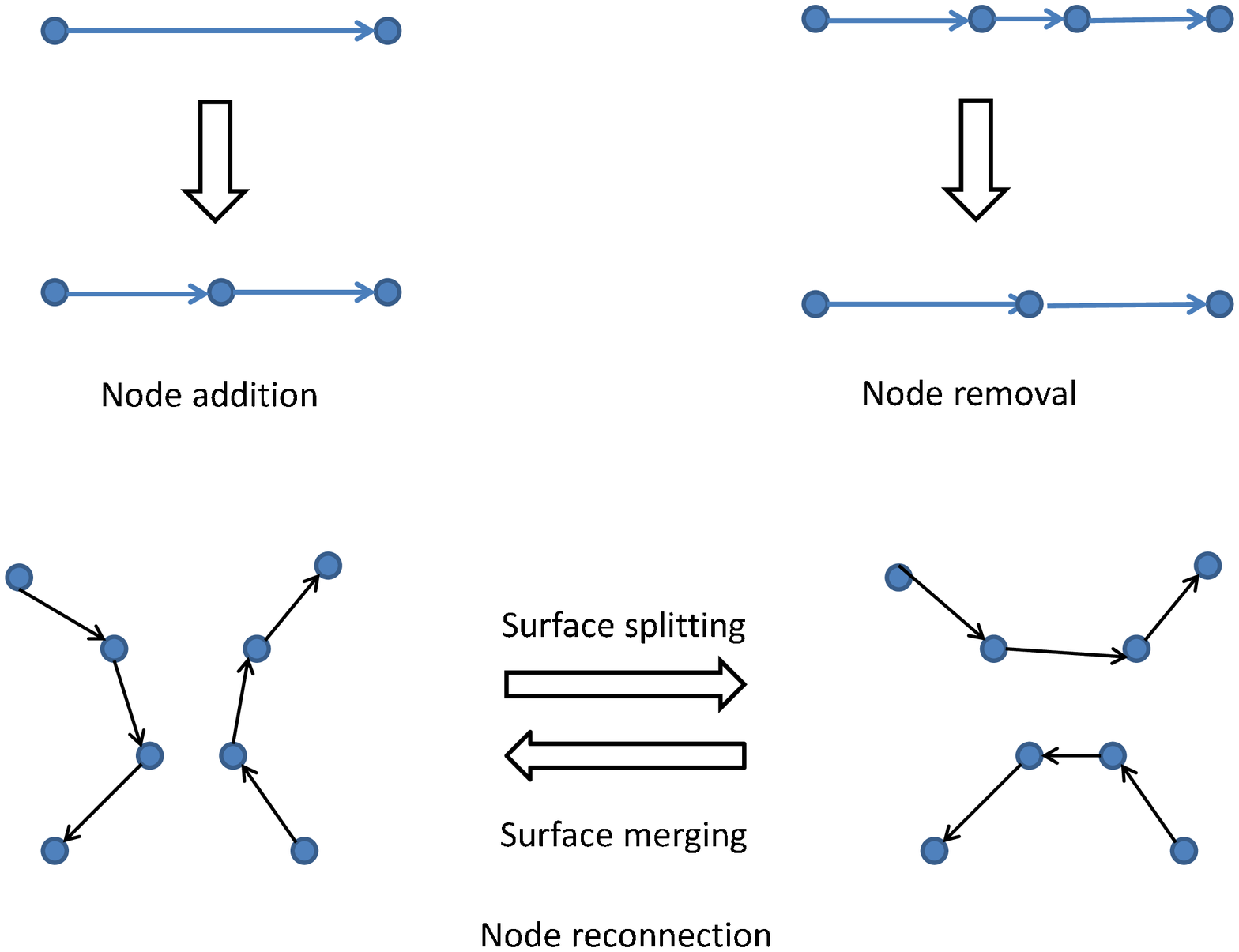}
\caption{Illustration on node addition, 
removal and reconnection. The arrows between nodes
stand for the pointer from one node variable 
to the next node variable.}
\label{fig:node_mani}
\end{figure}

To check whether there are regions
which are overcrowded or underpopulated with nodes, we 
calculate the distance between nearest neighbors. 
Given two nodes $i$ and $i+1$ with separation $d_{i,i+1}$, if 
$d_{i,i+1} < l_{{\rm min}}$, one of the node
is moved to a new position $((x_i+x_{i+1})/2, (y_i+y_{i+1})/2)$,
while the other node is deleted; on the 
other hand, if $d_{i,i+1} > l_{{\rm max}}$ 
an extra node between node $i$ and $i+1$ is inserted
at position $((x_i+x_{i+1})/2, (y_i+y_{i+1})/2)$. 
See Fig. \ref{fig:node_mani} for a graphical illustration of 
the above operations.

\begin{figure}
\centering
\includegraphics[width=8cm, height=6cm]{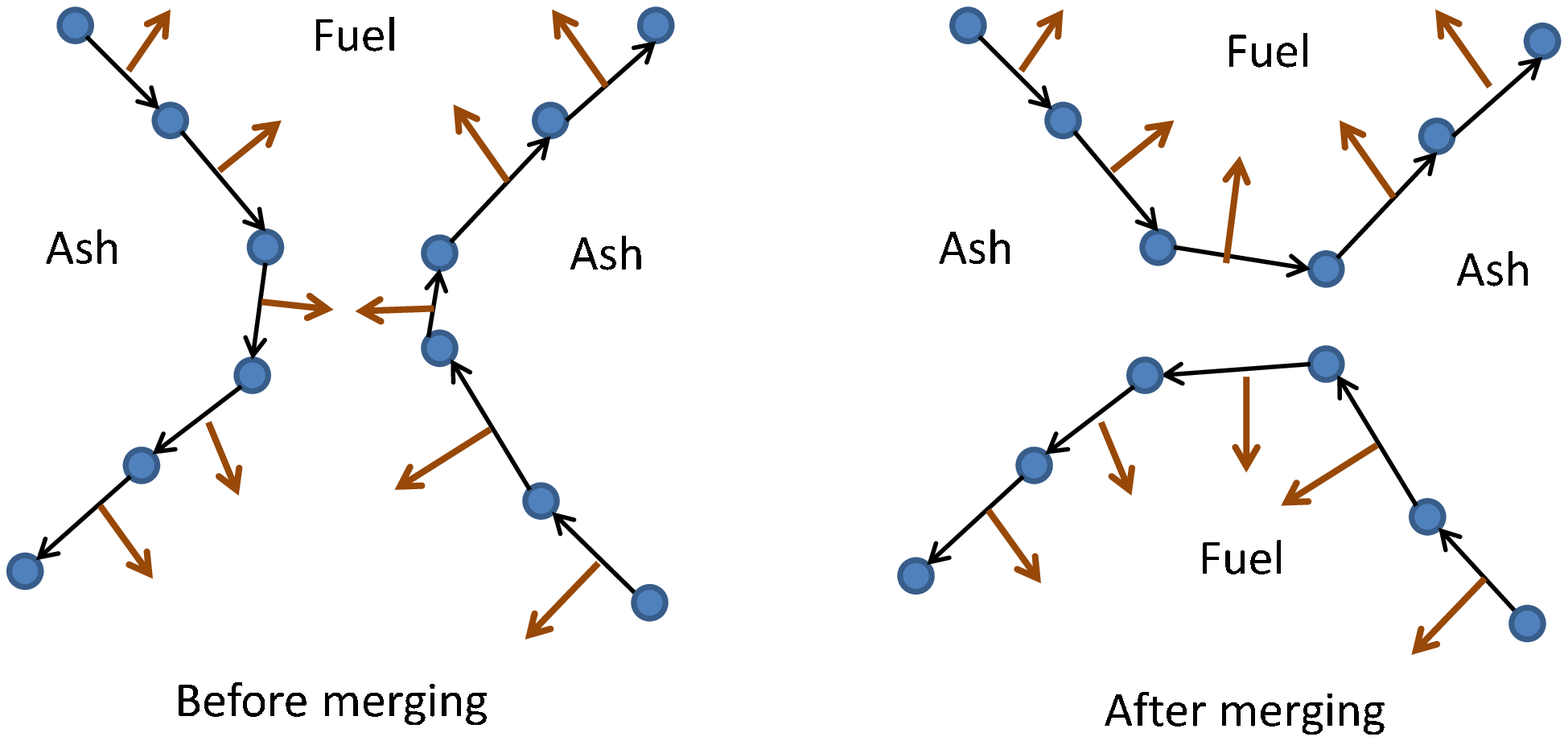}
\caption{Illustration on how node merging can
pervent two lines from crossing each other. The 
arrows on the line stand for the normal direction
of the flame surface.}
\label{fig:node_merge}
\end{figure}

To avoid lines from crossing each other, 
we locate all node pairs which are not connected
but are potentially close enough to form a line.
We first identify the nodes which lie on the 
same or neighboring grids, based on the same Eulerian
grid of the hydrodynamics, then the separations 
of this group of nodes are computed. When any pair
of non-neighboring nodes satisfies $d_{i,j} < l_{{\rm merge}}$
for some $i$ and $j$ that $|i - j| \neq 1$,
we reconnect the nodes as shown in Fig. \ref{fig:node_mani}
and change the topology of the surface. In principle, the aim is 
that the new surface may preserve the node position but the new 
surfaces will propagate away from each other, so that no 
entanglement can be formed in the coming time steps. 
In practice, for a line without entanglement, there exists
only one unpaired node $j$ with any node $i$ such that 
$d_{i,j} \leq l_{{\rm max}}$. 
However, when surfaces are going to split or merge,
there is more than one of such candidate,
which means that there is more than one way to form 
a line. To decide which node should be chosen,
we use the above principle to connect
the node such that a concave surface is obtained. 
Geometrically, the normal vectors on 
the new surfaces are converging. 
See Fig. \ref{fig:node_merge} for a graphical
description. We remark that how the surfaces merge is unique in 
two-dimensional simulations because all nodes have 
at most two neighbors for forming line segments. This
property is not true for three-dimensional models because
the surface is usually represented by triangular
patches, which means that each node always has
multiple connections with neighboring nodes.
Thus, the geometry of the surface, such as the 
curvature, will depend on how nodes are connected.

The three parameters, $l_{{\rm max}}$, $l_{{\rm min}}$
and $l_{{\rm merge}}$ are interrelated. 
First, we require $l_{{\rm max}} = 2 l_{{\rm min}}$. 
This is because we do not want any repetitive 
loop of node addition or removal to take place,
that means, after adding (removing) a new node between 
any two nearest-neighbors, the new configuration does not 
have nodes which are too close to (far from) 
each other according to the criteria. 
Similarly, we require $l_{{\rm merge}} \leq l_{{\rm max}}$ 
to avoid entangled lines in simulations. Since the point-set
method has a time step that prevents a node from moving further
than $l_{{\rm max}}$, we thus choose $l_{{\rm merge}} = l_{{\rm max}}$.

We compare the performance of the level-set algorithm 
with the point-set algorithm. The level-set algorithm
is known to be unable to track the flame
consistently when the flame propagates
much slower than the fluid flow. 
The flame surface is dominated by the turbulent
flow where the flame is supposed to show a convoluted
and elongated structure shaped by the fluid. 
But the level-set method can only preserve
structure with a size greater than the grid size. 
Therefore, the small-scale structure cannot be tracked
and a detached flame in forms of bubbles is obtained. To maintain
the consistency of the level-set algorithm in SNIa 
simulations, flame acceleration schemes are needed \citep{Calder2007}. 
To address this problem, we compare the flame structure
in the laminar flame limit by using the two algorithms. 
We present the simulation models in 
Table \ref{table:TestLine}. The hydrodyanmics
is the same as those in PTD tests and DDT tests. 
The hydrodynamics is done with a configuration similar to Section
\ref{sec:results_PTD} with an array of $500 \times 500$
in cylindrical coordinates with uniform grid size $\Delta = 11$ km.

\begin{table*}
\begin{center}
\caption{Simulation setup for the test of flame
capturing scheme. Lengths are in unit of 
code unit and densities are in unit of 
$10^{9}$ g cm$^{-3}$.}

\begin{tabular}{|c|c|c|c|c|c|}
\hline
Model & scheme & $l_{{\rm max}}$ & $l_{{\rm min}}$ & 
$l_{{\rm merge}}$ & $\rho_c$  \\ \hline
LSM-1 & level-set & 3.0 & 1.5 & 3.0 & 3 \\
PSM-1-2 & point-set & 2.0 & 1.0 & 2.0 & 3 \\
PSM-1-3 & point-set & 3.0 & 1.5 & 3.0 & 3 \\
PSM-1-4 & point-set & 4.0 & 2.0 & 4.0 & 3 \\ \hline
\label{table:TestLine}
\end{tabular}
\end{center}

\end{table*}

\begin{figure}
\centering
\includegraphics*[width=8cm, height=6cm]{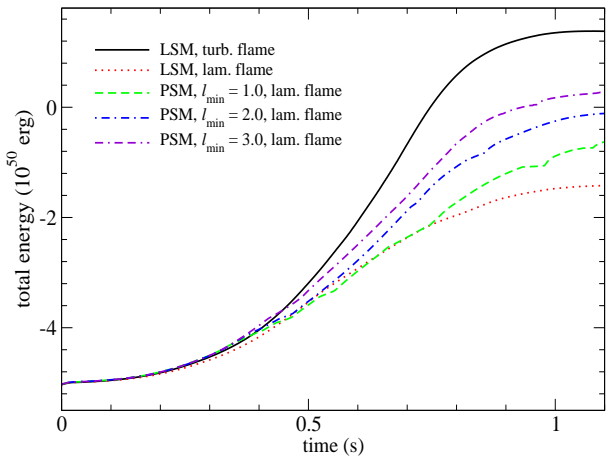}
\caption{Total energy against time for Models
LSM-1, PSM-1-2, PSM-1-3, PSM-1-4. See Table
\ref{table:TestLine} for the configurations.}
\label{fig:front_energy}
\end{figure}

In Fig. \ref{fig:front_energy} we plot the total energy 
against time for Models LSM-1, PSM-1-2, PSM-1-3
and PSM-1-4. As expected, the laminar deflagration
cannot successfully unbind the star. All 
four models perform similarly in early time.
The model PSM-1-3 is the most similar
one to LSM-1. At later time, the level-set method 
predicts less energy as compared to other three models.

\begin{figure}
\centering
\includegraphics*[width=8cm, height=6cm]{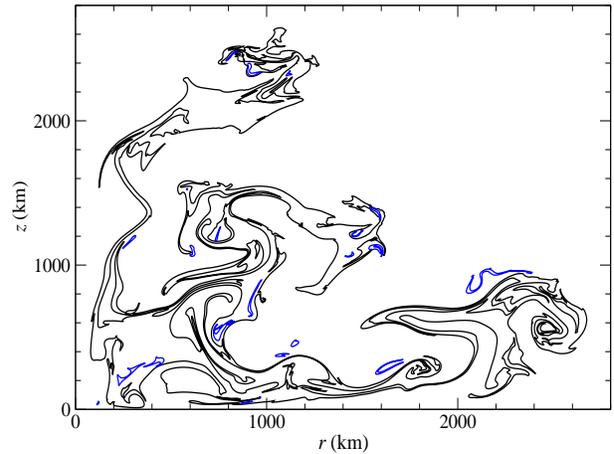}
\caption{Flame front of Model PSM-1-2
at $t = 1.00$ s.}
\label{fig:PSM-1-2-8}
\end{figure}

\begin{figure}
\centering
\includegraphics*[width=8cm, height=6cm]{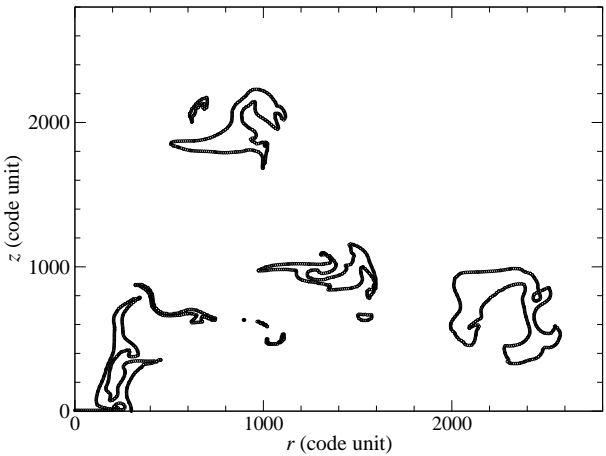}
\caption{Flame front of Model LSM-1
at $t = 1.00$ s.}
\label{fig:LSM-1-8}
\end{figure}

In Figs. \ref{fig:PSM-1-2-8}, \ref{fig:PSM-1-3-8}
\ref{fig:PSM-1-4-8} and \ref{fig:LSM-1-8} we plot the flame front
of Models PSM-1-2, PSM-1-3, PSM-1-4 and LSM-1, 
respectively at $t = 1.0$ s. By comparing 
Models PSM-1-2, PSM-1-3 and PSM-1-4, it shows
that the resolution of point-set method affects
the performance in two ways: First, it 
contributes to the fine details of RT and KH instabilities,
which enlarge the local flame area; second, the algorithm
allows an extremely elongated flame structure and an injection
of fuel into burnt region in the sub-grid scale. 
This property is important
for keeping information of the flame surface as much as 
possible, which maintains the fuel burning rate. 

\begin{figure}
\centering
\includegraphics*[width=8cm, height=6cm]{figC5.eps}
\caption{Flame front of Model PSM-1-3
at $t = 1.00$ s.}
\label{fig:PSM-1-3-8}
\end{figure}

\begin{figure}
\centering
\includegraphics*[width=8cm, height=6cm]{figC6.eps}
\caption{Flame front of Model PSM-1-4
at $t = 1.00$ s.}
\label{fig:PSM-1-4-8}
\end{figure}

We compare the point-set method and the 
level-set method by comparing the flame
surface of LSM-1 and PSM-1-4. Both methods
give thin flame shape. The point-set method 
shows a largely connected structure with some fuel
regions surrounded by ash. In level-set method, the flame is
broken into pieces and the flame bubbles are disassociated and 
are apart from each other. There is no 
sign of fuel injection. The upper part
of the flame can be compared with Model PSM-1-2. 
But the narrow parts in Model PSM-1-2
which extends from the core are not seen in LSM-1. 
Only the largest structure above the resolution size is preserved.
Comparing these results, the point-set
method can successfully capture the flame structure
even in the laminar flame limit, which is essential in 
understanding the energy release rate of highly 
convoluted flame.

\section{Ionization and Opacities}
\label{sec:opa}

In Section \ref{sec:method_radtran2} we described a 
hydrodynamics scheme in modeling the evolution 
of radiation of matter for a few weeks
after the SNIa explosion. In general, due to 
the rapid expansion and radiation, 
the matter reaches a temperature
where the assumption of complete ionization 
becomes invalid. Therefore, including ionization 
fraction in the light curve modeling is important for 
a consistent description of the thermodynamics
properties of the matter.

We use an open-source Saha-equation solver
\footnote{Provided as open-source code
in www.cococubed.edu. Refer \citep{Paxton2010, Paxton2013} for 
its applications in stellar evolution.}.
The subroutine solves the Saha 
equations, which relate the number 
densities of the $i^{{\rm th}}$-times ionized and the 
$i+1$-times ionized species by
\begin{equation}
\frac{n_{i+1,Z} n_e}{n_{i,Z}} = \Phi_{i,j,T_R}, 
\end{equation}  
where $\Phi_{i,Z}$ is a function of the 
partition functions of both species
at a given temperature and ionization
energies. In the subroutine, 
elements from $^{1}$H to $^{30}$Zn are included
and all ionization stages are considered.
An initial guess is given and then 
the solution is obtained by iterations.  

The ionization fractions of all elements
are then summed to find the number density 
of free electrons, which is used by the Helmholtz 
EOS subroutine in order 
to solve for the hydrodynamics pressure, 
internal energy and other local thermodynamics 
quantities. 

After the number densities of free electron $n_e$
and the number fraction $n_{Z,i}$ of an element $Z$
in an ionization stage $i$ are found, we obtain
the total opacity $\kappa$, which includes the Thomson
opacity $\kappa_e$, bound-free opacity $\kappa^{bf}(\nu)$
and free-free opacity $\kappa^{bb}(\nu)$
\begin{equation}
\kappa(\nu) = \kappa_e + \kappa^{bf}(\nu) + \kappa^{bb}(\nu).
\end{equation}
The Thomson opacity is given by
\begin{equation}
\kappa_e = \frac{n_e \sigma_e}{\rho}
\end{equation}
where $\sigma_e \approx 6.65 \times 10^{-25}$ cm$^{2}$ 
is the Thomson cross-section of electron.
The Thomson opacity is temperature
and density independent to a good approximation.

The bound-free opacity, or the 
photo-ionization opacity, is given by
\begin{equation}
\kappa^{bf}(\nu) = \sum_{Z,i} \frac{\sigma^{bf}_{Z,i} (\nu) n_{Z,i}}{\rho},
\end{equation}
with $\sigma^{bf}_{Z,i} (\nu)$ being the 
frequency-dependent bound-free scattering
cross-section for the element $Z$ at the $i^{{\rm th}}$
ionization stage. We use the fitting formula 
reported by \cite{Verner1996}
\begin{equation}
\sigma^{bf}_{Z,i}(\nu) = \sigma_0 \left[ (x(\nu)-1)^2 + y_W \right] 
y^{P/2 - 5.5} (1 + \sqrt{\frac{y}{y_a}})^{-P},
\end{equation}
where 
\begin{equation}
x(\nu) = \frac{\nu}{e_0} - y_0
\end{equation}
and 
\begin{equation}
y = \sqrt{x^2 + y^2_1}.
\end{equation}
The constants $e_0$, $\sigma_0$, $y_0$, $y_1$, $y_a$, 
$y_w$ and $P$ are constants depending on 
the species and ionization stage.

The free-free opacity is given by \citep{Sakamoto2001}
\begin{equation}
\kappa^{ff}(\nu) = \sum_{Z,i} C_{Z,i} n_e n_{Z,i} \rho \sqrt{T} \nu^{-3},
\end{equation}
with $C_{Z,i}$ being some constants related to the
Gaunt factor, whose values are fitted in 
\citep{Itoh1986, Nozawa1998, Itoh2000}.

The Rosseland mean opacity is then given by
\begin{equation}
\frac{1}{\kappa_R} = \frac{\int^{\infty}_0 \frac{1}{\kappa(\nu)} 
\frac{dB_{\nu} (T)}{dT} d\nu}{\int^{\infty}_0 \frac{dB_{\nu}(T)}{dT} d\nu},
\end{equation}
with $B_{\nu}(T)$ being the Planck distribution function. 
Since only the bolometric light curve is modeled, 
we assume $\kappa_P = \kappa_E$ and $\chi_F = \chi_R$
in the calculation. 

The effects of lines are important because of the 
Doppler effect inside the fast expanding ejecta.
Photons with frequencies within 
Doppler widths from the atomic scattering lines
can be absorbed or scattered. 
To include the line opacity, we follow the prescription
in \cite{Karp1977} and \cite{Hoeflich1993}. We
first compute the Sobolev opacity of each line by 
\begin{equation}
\kappa_{{\rm line}}(i,j) = \frac{\pi e^2}{m_e c} \frac{f_{ij} n_i}{\nu_{ij} \rho} 
\left( 1 - \frac{g_i n_j}{g_j n_i} \right),
\end{equation}
with $f_{ij}$ being the oscillator strength
between atomic states $i$ and $j$, $n_i$ and $g_i$
are the occupation number 
and statistical weight at state $i$. 
$\nu_{ij}$ is the transition frequency.

The effective absorption coefficient $\chi_{\nu}$
is obtained by summing all relevant lines, 
where the lines are arranged in increasing wavelength 
\begin{eqnarray}
\chi_{\nu} && = \kappa_{{\rm cont}} \times \left[ 1 - \right.  \nonumber \\
&& \left. \sum^N_{j=J} 1 - {\rm exp}(- \tau_j)
\left( \frac{\nu_j}{\nu} \right)^{s (\nu)} {\rm exp} 
\left( - \sum^{j-1}_{i=1} \tau_i \right) \right]^{-1},
\label{eq:lineopc}
\end{eqnarray}
with 
\begin{equation}
\tau_i = \kappa_{{\rm line}} c \rho \frac{dr}{dv(r)}
\end{equation}
being the effective optical depth of the line $i$. 
$\kappa_{{\rm cont}}$ is the continuum opacity
(bound-free opacity, free-free opacity
and Thomson scattering). The frequency dependent 
expansion rate parameter $s(\nu)$ is given by
\begin{equation}
s(\nu) = \kappa_{{\rm cont}} c \rho \left( \frac{dv}{dr} \right)^{-1}.
\end{equation}
We remark that as pointed out in \cite{Blinnikov1996} 
and \cite{Pinto2000b}, the summation in Eq. \ref{eq:lineopc}
should be done in an upwind direction so that the opacity 
includes only the lines that are being scattered.

To relate the extinction coefficient with the absorption coefficient,
we define the frequency independent enhancement factor 
\begin{equation}
\epsilon = \frac{\chi_R - \sigma_e}{\chi_R},
\end{equation}
with $\chi_R$ the same Rosseland opacity but with 
the line opacity included. To discriminate the 
difference between line absorption and line scattering, 
we define the ratio between line collisional and 
radiative transition rate between state $i$ and 
$j$ by $\alpha$, which modifies the the 
enhancement factor by an extra factor
\begin{equation}
\epsilon' = \frac{\alpha}{\alpha + 1} \epsilon.
\end{equation}
The frequency dependent absorption coefficient is 
\begin{equation}
\kappa_{\nu} = \epsilon' (\chi_{\nu} - \sigma_e).
\end{equation}

\end{document}